\documentclass[a4paper,twoside,11pt]{article}
\usepackage{a4}
\usepackage{graphics}

\begin{document}
%%%%%% text sizes etc %%%%%%%%%%%%%%%%%%%%%
\textheight23cm
\textwidth15.5cm
\oddsidemargin0.5cm
\evensidemargin0cm
\parskip1ex
\pagestyle{myheadings}
%%%%%%%%%%%%%%%%%%%%%%%%%%%%%%%%%%%%%%%%%%%

%%%%%%%%%%%%%%%%%%%%%%%%%%%%%%%
\def\sun{\hbox{$\odot$}}
\def\la{\mathrel{\mathchoice {\vcenter{\offinterlineskip\halign{\hfil
$\displaystyle##$\hfil\cr<\cr\sim\cr}}}
{\vcenter{\offinterlineskip\halign{\hfil$\textstyle##$\hfil\cr
<\cr\sim\cr}}}
{\vcenter{\offinterlineskip\halign{\hfil$\scriptstyle##$\hfil\cr
<\cr\sim\cr}}}
{\vcenter{\offinterlineskip\halign{\hfil$\scriptscriptstyle##$\hfil\cr
<\cr\sim\cr}}}}}
\def\ga{\mathrel{\mathchoice {\vcenter{\offinterlineskip\halign{\hfil
$\displaystyle##$\hfil\cr>\cr\sim\cr}}}
{\vcenter{\offinterlineskip\halign{\hfil$\textstyle##$\hfil\cr
>\cr\sim\cr}}}
{\vcenter{\offinterlineskip\halign{\hfil$\scriptstyle##$\hfil\cr
>\cr\sim\cr}}}
{\vcenter{\offinterlineskip\halign{\hfil$\scriptscriptstyle##$\hfil\cr
>\cr\sim\cr}}}}}
\def\degr{\hbox{$^\circ$}}
\def\arcmin{\hbox{$^\prime$}}
\def\arcsec{\hbox{$^{\prime\prime}$}}
\def\utw{\smash{\rlap{\lower5pt\hbox{$\sim$}}}}
\def\udtw{\smash{\rlap{\lower6pt\hbox{$\approx$}}}}
\def\fd{\hbox{$.\!\!^{\rm d}$}}
\def\fh{\hbox{$.\!\!^{\rm h}$}}
\def\fm{\hbox{$.\!\!^{\rm m}$}}
\def\fs{\hbox{$.\!\!^{\rm s}$}}
\def\fdg{\hbox{$.\!\!^\circ$}}
\def\farcm{\hbox{$.\mkern-4mu^\prime$}}
\def\farcs{\hbox{$.\!\!^{\prime\prime}$}}
\def\fp{\hbox{$.\!\!^{\scriptscriptstyle\rm p}$}}
%%%%%%%%%%%%%%%%%%%%%%%%%%%%%%%%%%%%%%%%%%%%%%%%%%%%%%%%%%%%%%%%%%%%%%

\newcommand{\D}{\displaystyle} %normal formulas
\newcommand{\T}{\textstyle} %for large font
\newcommand{\SC}{\scriptstyle} %footnote
\newcommand{\SSC}{\scriptscriptstyle} %footnote to footnote

\def\AJ{{\it Astron. J.} }
\def\ARAA{{\it Annual Rev. of Astron. \& Astrophys.} }
\def\ApJ{{\it Astrophys. J.} }
\def\ApJL{{\it Astrophys. J. Letters} }
\def\ApJS{{\it Astrophys. J. Suppl.} }
\def\ApP{{\it Astropart. Phys.} }
\def\AA{{\it Astron. \& Astroph.} }
\def\AAR{{\it Astron. \& Astroph. Rev.} }
\def\AAL{{\it Astron. \& Astroph. Letters} }
\def\JGR{{\it Journ. of Geophys. Res.}}
\def\JHEP{{\it Journal of High Energy Physics} }
\def\JPhG{{\it Journ. of Physics} {\bf G} }
\def\PhFl{{\it Phys. of Fluids} }
\def\PASJ{{\it Publ. Astron. Soc. Japan} }
\def\PR{{\it Phys. Rev.} }
\def\PRD{{\it Phys. Rev.} {\bf D} }
\def\PRL{{\it Phys. Rev. Letters} }
\def\Nature{{\it Nature} }
\def\MNRAS{{\it Month. Not. Roy. Astr. Soc.} }
\def\ZA{{\it Zeitschr. f{\"u}r Astrophys.} }
\def\ZFN{{\it Zeitschr. f{\"u}r Naturforsch.} }
\def\etal{{\it et al.}}

\hyphenation{mono-chro-matic  sour-ces  Wein-berg
chang-es Strah-lung dis-tri-bu-tion com-po-si-tion elec-tro-mag-ne-tic
ex-tra-galactic ap-prox-i-ma-tion nu-cle-o-syn-the-sis re-spec-tive-ly
su-per-nova su-per-novae su-per-nova-shocks con-vec-tive down-wards
es-ti-ma-ted frag-ments grav-i-ta-tion-al-ly el-e-ments me-di-um
ob-ser-va-tions tur-bul-ence sec-ond-ary in-ter-action
in-ter-stellar spall-ation ar-gu-ment de-pen-dence sig-nif-i-cant-ly
in-flu-enc-ed par-ti-cle sim-plic-i-ty nu-cle-ar smash-es iso-topes
in-ject-ed in-di-vid-u-al nor-mal-iza-tion lon-ger con-stant
sta-tion-ary sta-tion-ar-i-ty spec-trum pro-por-tion-al cos-mic
re-turn ob-ser-va-tion-al es-ti-mate switch-over grav-i-ta-tion-al
super-galactic com-po-nent com-po-nents prob-a-bly cos-mo-log-ical-ly
Kron-berg Rech-nun-gen La-dungs-trenn-ung ins-be-son-dere
Mag-net-fel-der bro-deln-de}
%Formeln
\def\simle{\lower 2pt \hbox {$\buildrel < \over {\scriptstyle \sim }$}}
\def\simge{\lower 2pt \hbox {$\buildrel > \over {\scriptstyle \sim }$}}

% instructions for automatic equation numbering
%Benutzung:  $$ Formel\eqno\autnum$$

\begin{center}
{{\bf Cosmic Rays from PeV to ZeV, Stellar Evolution,
Supernova Physics and Gamma Ray Bursts}}\\
\vskip1.0cm
Peter L. Biermann$^{1,2}$, Sergej Moiseenko$^{3,1}$,
Samvel Ter-Antonyan$^{4,1}$, \& Ana Vasile$^{1,5}$\\[10mm]
$^1$ Max-Planck Institut f{\"u}r Radioastronomie, Bonn, Germany\\
$^2$ Department for Physics and Astronomy, University of
Bonn\\
$^3$ Space Research Institute, Moscow, Russia \\
$^4$ Yerevan Physics Institute, Erewan, Armenia\\
$^5$ Institute for Space Sciences, Bucuresti-Magurele, Romania\\
plbiermann@mpifr-bonn.mpg.de,moiseenko@mx.iki.rssi.ru,
samvel@jerewan1.yerphi.am,avasile@venus.nipne.ro\\[10mm]
www.mpifr-bonn.mpg.de\\/div/theory/ \\
www.mpifr-bonn.mpg.de/imprs/ \\
www.physik-astro.uni-bonn.de
\end{center}

%  moiseenko@mx.iki.rssi.ru,samvel@jerewan1.yerphi.am,\\
%  anavasile@hotmail.com,avasile@venus.nipne.ro,
%  alternate email:       Samvel <samuel@web.am>
%
%  maybe consult:
%  Sabrina Casanova, Ralph Engel, Norbert Langer, Athina Meli, ... \\
%  casanova,Ralph.Engel@ik.fzk.de,N.Langer@astro.uu.nl,a.meli@ic.ac.uk,

%  also check with Eun-Suk Seo, and Todor Stanev \\
%  es83@umail.umd.edu,stanev@bartol.udel.edu

% -----------------------------
% final version Feb 4, 2003, final revision Feb 9, 2003
%------------------------------

\section{Abstract}

The recent success of a proposal from some time ago to explain the
spectrum of cosmic rays allows some strong conclusions to be made on the
physics of supernovae:  In the context of this specific proposal to
explain the origin of cosmic rays, the mechanism for exploding supernovae
of high mass has to be the one proposed by Bisnovatyi-Kogan more than 30
years ago, which was then based on a broader suggestion by Kardashev:  A
combination of the effects of rotation and magnetic fields explodes the
star.  Interestingly, this step then leads inevitably to some further
suggestions, useful perhaps for the study of gamma ray bursts
and the search of a bright standard candle in cosmology.

\section{Introduction}

Cosmic Ray physics has inspired us for many years, and has given rise to
many interesting books and reviews, some of which are
\cite{GS63,Venyabook,Gaisser90,Bhatta99}.

Several burning questions exist in high energy astrophysics:

\begin{itemize}

\item{}  What is the physics of the supernova explosion of massive
stars?  The most common idea to be explored over the last few decades is
that the burst of neutrinos, certainly sufficient in energy, as shown by
the supernova 1987A, is doing it.  However, the detailed mechanism has
yet to be worked out successfully.

\item{}  What is the origin of Galactic cosmic rays? What makes it
possible for cosmic rays from Galactic sources to reach energies such as
$3 \, 10^{18}$ eV?

\item{}  What is the mechanism of Gamma Ray Bursts, and what is it's
relation to supernovae?

\item{}  Is there a brighter standard candle possibly available for
cosmology than the now so famous Supernovae type Ia?

\end{itemize}

Here, in this review, we wish to combine several recent advances made to
give a tentative answer to all these questions, starting with the work on
cosmic rays:

\section{The cosmic ray spectrum}

We wish to explain the entire spectrum of cosmic rays, and focus here on
Galactic cosmic rays, so the energies up to about $3 \, 10^9$ GeV.  Some
time ago we made a proposal to explain Galactic and extragalactic
cosmic rays in six consecutive papers,
\cite{CRI,CRII,CRIII,CRIV,UHECRI,UHECRII}.  The basic idea was to
distinguish the different sites of supernovae into the lower
mass stars, that explode into the interstellar medium, and the higher
mass stars that explode into their own stellar wind.  The basic
data to explain are as follows, \cite{LB-CR,NW2000}:

\begin{itemize}

\item{}  The overall spectrum is about approx. $E^{-2.7}$ until the
``knee", which is a bend downwards at around $3 \, 10^{15}$ eV

\item{}  approx. $E^{-3.1}$ beyond the knee

\item{}  a slight downward dip from $3 \, 10^{17}$ eV, sometimes referred
to as the ``second knee"

\item{}  a transition near $3 \, 10^{18}$ eV, with then approx.
$E^{-2.7}$ again

\item{}  uncertainty beyond $5 \, 10^{19}$ eV, either a mild cutoff
(HIRES) or a continuation (AGASA)

\end{itemize}

and in more detail, considering electrons and nuclei:

\begin{itemize}

\item{}  electrons $E^{-2.7}$ as well (from radio, low $E$), up to about
20 GeV

\item{}  electrons $E^{-3.3}$ observed to 3 TeV: \\
loss dominated, so injection $E^{-2.3}$

\item{}  positron fraction a few percent

\item{}  abundances enriched

\item{}  anti-proton fraction about $10^{-4}$

\item{}  heavy elements tend to have a slightly flatter spectrum

\end{itemize}

\section{The arguments}

Already in 1934 the original suggestion by W. Baade \& F. Zwicky was that
the most energetic particles probably come from the most energetic
phenomonea known to us at the time, supernovae, \cite{BZ34}.  E. Fermi
then provided an argument on how reflection at magnetic irregularities can
enhance the energy of a charged particle, 1949 and 1954,
\cite{Fermi49,Fermi54}.  The detailed physics of the Fermi-acceleration
was worked out in a simple approximation in a series of papers, starting
in 1977 by I. Axford \etal, G. Krymsky, \cite{Kr77}, R. Bell in two
papers, \cite{Bell78}, and then again by R. Blandford \& J. Ostriker,
\cite{BO78} with a nice extensive review by L. Drury, \cite{Drury83}.

\subsection{Supernova explosions into the ISM}

Then P. Lagage \& C. Cesarsky, \cite{Lagage+C83}, worked out the details
for the maximum energy of a particle subject to Fermi acceleration in the
shock of a supernova, exploding into the interstellar medium.
Presciently, they already used the self-similar expansion into a very
tenuous medium, now known to exist from X-ray observations.  For the
details of the acceleration of energetic particles at the shock they use
the concept that most of the acceleration happens in a regime, where the
magnetic field is parallel to the shock normal - as well its disturbances
- and provides most of the scattering.  Here the expansion solution is a
similarity or Sedov-solution, a self-similarity, where the outer shock
radius $r$ behaves as $r \sim t^{2/5}$, and the shock speed is $\dot{r}
\sim r^{-3/2}$, and so slows down considerably over time.  This slow late
evolution makes these supernova remnants dominant in observations, they
just live a rather long time.

\begin{itemize}

\item{}  $E_{max} \, \simeq \, 10^5 \, Z B_{-6} GeV$,
using already the hot ISM of density $3 \, 10^{-3} \, {\rm cm}^{-3}$,
confirmed only in 1997, \cite{ROSAT97}.  The expansion is self-similar,
usually referred to as a Sedov expansion, and yields then a maximum
particle energy of about $Z \, 10^{14}$ eV, an order of magnitude below
the knee.

\item{}   Maybe secondary acceleration is then required in the ISM - maybe
by shockwaves inside superbubbles, the motley mixture of powerful stellar
winds, and young supernova explosions.  This picture has mostly been
explored by I. Axford.  A quantitative prediction is still outstanding
for this scenario.

\end{itemize}

\subsection{The knee}

The key problem, recognized immediately, is that this maximum energy does
not even come to the energy at the knee, and so that feature cannot be
explained in such an approach.  Therefore many attempts have been made:

\subsubsection{Transport?}

Cosmic rays get injected with a spectrum, which is subsequently modified
by losses from the disk, and so obviously, to get a kink in the spectrum
may mean a change in behaviour of these diffusive losses; such a kink
could be visible as the ``knee".  This is a question of the transport
of cosmic rays through a turbulent magnetic ionized gas;
general discussions of the magnetic field are in, e.g.,
\cite{Kronberg94,Kulsrud99}.   Such a change in transport implies a
special scale in the interstellar medium, and would also lead to an
increased anisotropy at the kink.  However, there is no such scale as yet
recognized.

\subsubsection{Turbulence?}

The evidence from the interstellar medium and from plasma physics
suggests - if described by an isotropic approximation with an eye towards
transport of cosmic rays - that the spectrum is a Kolmogorov
spectrum.  There are three arguments in favour of using a Kolmogorov
spectrum for the context of cosmic rays: First, the difference in
behaviour between nuclei, and electrons, allows the exponent of the
diffusive law to be limited by about 0.4, so the energy dependence of the
diffusion coefficient $\kappa$ is weaker than
$\kappa \sim E^{0.4}$, \cite{TucsonCR}.  Second, any steeper dependence
would produce anisotropies at the higher energies, which are not seen,
\cite{CRI}.  And, finally, any steeper dependence would seriously dilute
the number of stars contributing at the higher energies, making any
reasonable powerlaw fit to the data meaningless, despite the fact, that
such fits work well to describe the data, \cite{VulcanoCRa,VulcanoCRb}.

\subsubsection{Spallation?}

Cosmic rays interact, and in fact most isotopes of Lithium, Beryllium, and
Bor, as well as the odd-Z elements, and the sub-iron elements arise from
cosmic ray interaction, the destruction of nuclei in a spallation process,
see M. Garcia-Munoz \etal, \cite{GM77,GM87}.   Therefore, a change in this
interaction may lead to a loss of nuclei at higher energy; the
interstellar medium is discussed, with some emphasis on the Galactic
Center region, e.g., in \cite{Mezger96}.  However, when working through
this idea, one finds that the column for interaction must be much higher
than what is observed; in fact, from the abundances of all the spallation
products, as well as positrons and anti-protons we have a fairly good
idea what the interaction actually is, and it is not enough to provide
such a drastic change.  Furthermore, that data suggest today, that the
``knee" is a feature in rigidity $E/Z$, and so plasma physical processes
are implied.

\subsubsection{New source?}

Obviously, in some proposals we require a new source at $3 \, 10^{18}$ eV,
or maybe even at $10^{14}$ eV, and so one might realistically ask,
whether a new source might not be necessary already at lower energies, if
that source is supposed to work so well at higher energies.  A new source
is easiest to accomodate, if the new source, dominant at higher energy,
has a flatter spectrum than the source dominant at lower energy; in such
a case the transition is trivial, since the higher energy source just
overtakes the lower energy source naturally in flux.  However, if the
transition is supposed to occur at or near the knee, then the higher
energy source must have a steeper spectrum rather than a flatter spectrum,
and so we require a multitude of strange coincidences:  The higher energy
source must have a peaked spectrum, peaking just when the lower energy
source peters out, and so have the same flux at that energy, while not
contributing at lower energy in any significant way.  That is really hard
to believe.  So, it is then actually simpler, and has been suggested,
that we use a new source from the start, and, e.g., adopt the point of
view, that Gamma Ray Bursts produce all the cosmic rays, in work by C.
Dermer (discussed in a talk at the Aspen meeting, January 2002).  However,
then we are back where we started, we still need an argument why we have a
change in slope at a specific rigidity, the knee.  Also, the expected
flux of cosmic rays from Gamma Ray Bursts is nowhere near to what is
necessary, \cite{GRB1,GRB2}.  However, Gamma Ray Bursts should produce
cosmic rays, and the question is whether there is any way to ascertain
their contribution, perhaps through high energy neutrinos. Another
possible source could be the activity in the Galactic Center,
\cite{LM2000,MF2001}.  However, there is no observational evidence at this
time, that such a localized activity could be dominant at the location of
the Solar system.

\subsubsection{Change in acceleration?}

Therefore, what remains, is more mundane perhaps, a small modification in
acceleration of the existing cosmic rays.  The proposal by I. Axford
\etal  $\;$ is that normal supernovae provide cosmic rays to about the
knee, and that then in the supernova wind bubbles around the really
massive stars, the cosmic rays get further acceleration, and then have a
slightly diminished efficiency in acceleration; after all, the detailed
physical properties are somewhat different in the environment of many
supernovae and their predecessor winds, as compared to the interstellar
medium.  The only step missing in this very nice argument is a
quantitative prediction as to what expect in this picture; one may
surmise that the predictions may not turn out to be quite so different
from the proposal described below.

\section{Proposal for the Origin of Galactic Cosmic Rays}

The riddle, which provoked a new approach now some time ago, was the key
observation, that the radio shells of supernova remnants show a
radial direction for the magnetic field, parallel to the shock-normal.
This is in exact contradiction to the simple plasma physics expectation
that the enhancement of the magnetic field component parallel to the
shock surface should make that component dominant behind the shock, see
\cite{CRIII}.

Additional support for a new approach came from the observation that
radio features could be seen in motion, with an amplitude of velocity
quite close to the overall expansion speed of the supernova shock front,
and yet measured with respect to the expanding frame of reference.

Third, the thickness of the observed shells both in the radio and also to
some degrees in X-rays, is much too large to allow a simple explanation.
There must be some reason for these shells to be almost 1/3 of the outer
radius oberved.

Finally, there appeared a theoretical argument, that cosmic ray dominated
shocks are unstable, in work by R. Ratkiewicz, I. Axford, G. Webb, e.g.,
\cite{RAMK94}.

Therefore, we introduced the key premise that the particle transport in
the shock region is dominated by large scale turbulence; in such a case
there is no time for any turbulent cascade, and so no time for a
Kolmogorov cascade either.  Therefore the transport is just given by a
specific scale, and a specific velocity, naturally the thickness of the
shocked region, and the velocity difference across the shock.  This is in
the spirit of the early work on turbulence such as by L. Prandtl and
Th.v. Karman.  This premise allows then to explain all these observed
features.

This means that the shock region is strongly turbulent, and the front of
the shock, as seen along the line of sight at the edge of some observed
supernova remnannt, is actually a superposition of many smaller shocks,
advancing and retreating, like the waves on a beach.  The geometric scale
of these shocks is again given by the radial overall shock thickness,
just given by momentum conservation.  Therefore, curvature drifts do play
a role, and introduce a scale in energy per charge $E/Z$.

It can be surmised, that the limit cycle arguments by M. Malkov \& L.
O'C. Drury will eventually lead to a similar conclusion, \cite{MOD01}.
Such a convergence of theoretical concepts will be an important test of
the picture.

\subsection{Supernovae into a stellar wind}

The concept then leads to the consideration of supernovae exploding into
their own stellar wind, \cite{HJV+PLB88,CRI,CRII}.  Stellar winds may be
considered in a first approximation as a Parker wind, with
$B_{\phi} \sim \sin \theta / r$ in spherical coordinates.  The dominant
magnetic field character is an Archimedian spiral, with the field mostly
very nearly tangential.  The key aspects are i) that the wind is
magnetic, with the magnetic field given by the star, and not the
interstellar medium, and ii) that the wind has a density gradient
$\sim r^{-2}$, which means that the shock stays fast, and does not slow
down as in an explosion into a medium of constant density.  Also, magnetic
fields in massive stars exist, they have been detected through nonthermal
radio emission and maser emission, \cite{CRII}.

As the magnetic fields get very weak and also radial towards the pole of a
Parker-type wind, one needs to consider shock acceleration in two limiting
regimes:

At the pole and near to it we need to use the acceleration dominated by
the magnetic field parallel to the shock normal, as done by
\cite{Lagage+C83}.  However, at the equator, where the shock direction is
perpendicular to the dominant magnetic field, we need to use another
limit, as shown by R. Jokipii, \cite{Jokipii87}.  Therefore we also need a
matching condition, and this leads to a proposal for identifying the knee
as arising with this matching energy, the maximum energy near the pole
and close to it:

\begin{equation}
E_{max} \, = \, Z e r B (3/4 \; v_{sh}/c)^2  \label{eq:Em1}
\end{equation}

The maximum energy in the equatorial region, and in fact, most of $4 \pi$
is

\begin{equation}
E_{max} \, = \, Z e r B  \label{eq:Em2}
\end{equation}

Since $B$ is inversely proportional to radius $r$ in the wind, these
energies are actually constant throughout the wind, see e.g.
\cite{SB97}.
We still need to show, that the particles actually have enough time to
reach these maximum energies, but this appears plausible.  We also need to
show, why this leads to a different spectrum beyond the knee:  This is
due to the effect that curvature drifts, which provide a small proportion
of the particle's energy gain, when subject to many shock transitions, get
slightly weaker when the Larmor radius of the particle moves past the
scale corresponding to the maximum energy (really $E/Z$) near the pole,
see \cite{CalgaryCR} for a very detailed discussion.

Those stars which have powerful winds, are usually enriched in chemical
elements such as Helium, Carbon and Oxygen, and so this line of reasoning
actually picks up a theme, that Wolf Rayet stars are a major
contributor to cosmic rays, \cite{ST90}.  Therefore, supernova explosions
into stellar winds are proposed to provide most of the Galactic cosmic
rays, with the notable exception of the element Hydrogen, for which they
provide only some fraction.

\section{Stellar evolution}

For the physics of cosmic rays and supernovae we need to consider
stellar evolution in some detail.  What happens to stars of different
zero-age main sequence masses $M$:

\begin{itemize}

\item{}  Single stars with $M < \; 8 \; M_{\odot}$ give no supernovae.

\item{}  Stars in the mass range $8 \; < \; M \; < \; 15  M_{\odot}$
explode as supernovae, but the explosion goes into the interstellar
medium.  Such explosions lead a Sedov type expansion, which becomes slow
and has a long-lasting remnant.

\item{}  Stars with $15 \; < \; M \; < \; 30  M_{\odot}$ explode
as supernovae with a substantial wind, but
enriched only in Helium.  The mass in the wind-shell is moderate.  We
refer to these stars as Red Super Giant stars (RSG).

\item{}  Stars $> \; 30 \; M_{\odot}$ explode as supernovae into a strong
wind; the chemical abundances are strongly enriched, since the mass loss
eats deeply down back into the star.  The mass in the wind-shell is
large.  We refer to these stars as Wolf-Rayet stars (WR).

\end{itemize}

Obviously, these numbers for the masses are approximations.

\section{Predictions}

The predictions, all made some time ago, here focussing on wind-SNe, were
described in \cite{CRI,CRII,CRIII,CRIV,CalgaryCR,TucsonCR}.

They gave the following quantitative spectra:

\begin{itemize}

\item{}  $E^{-2.74 \pm 0.04}$ for the interstellar medium supernovae,
with the abundances that correspond to the medium through which the
supernova shock races.  This spectrum is predicted to have a cutoff in the
100 TeV range.

\item{}  $E^{-2.67 - 0.02 \pm 0.02}$ below knee for the wind supernovae,
for which the shock races through the predecessor wind, and
correspondingly has enriched abundances.

\item{}  $E^{-3.07 - 0.07 \pm 0.07}$ above knee for those same stars
winds.

\item{}  A bending of the wind spectra at the ``knee" near $Z \, 10^{15}$
eV, and a final cutoff at the ``ankle" near $Z \, 3 \, 10^{17}$ eV.  These
specific rigidities have an uncertainty of about a factor of 2.

\item{}  When the supernova explodes, a powerful shock wave races through
the wind, and then smashes into the shell; this wind-shell is made up of
both wind-material and interstellar medium material, from the environment
of the predecessor star.  Such a shell has two shocks at its boundaries,
one slowing down the wind, and one speeding up the interstellar
material.  If there is no convective instability, then a contact
discontinuity separates wind material from environmental interstellar
medium material.

\item{} The shock is loaded with energetic particles from cosmic ray
acceleration, and then these cosmic rays suffer from spallation in the
shell, \cite{HirscheggCR,CR9}.  Both cosmic ray particles as well as wind
material particles get broken up.  This, in the case of WR shells, gives
the Be, B, Li nuclei in cosmic rays, which serve then as tracers of
cosmic ray lifetimes, and interaction.  It also gives the odd-$Z$
elements, as well as the sub-Fe elements in cosmic rays.

\item{}  As a new prediction following from the work in \cite{CR9} we
find a small spectral flattening is expected due to differential
spallation for Fe-like nuclei.  The fraction of nuclei removed by
spallation from the set of Fe-like nuclei is so large, that across the
range of energy per particle measured, this fraction is larger for lower
energy, and so the spectrum flattened.  This proposal is quantitatively
consistent with the data.

\end{itemize}

At that stage we did not differentiate the cosmic ray acceleration
scenario for red supergiant winds (RSG), and Wolf Rayet (WR) winds.  We
now do differentiate those two kinds of stars with winds in their
interaction in the wind shell, \cite{CR9}.  The RSG stars have a weaker
wind, therefore a lower mass shell as compared to the WR stars, and so we
argued that the cosmic ray interaction, when the supernova shock finally
hits the shell, is convective for RSG stars, and diffusive for WR stars.
In this picture the RSG stars produce the gamma ray emission from the
Galactic plane, and the WR stars produce the spallation products such as
Beryllium, and give only some small contribution to the Galactic gamma ray
emission at moderate photon energies.

\subsection{Implications}

Such a model implies, as had been predicted already by B. Peters in the
fifties, \cite{Peters59,Peters61}, that the turnover at the knee, and at
the final cutoff is gradual, as the elements roll off in sequence of $Z$.
Furthermore, since the knee and final cutoff energy are given by stellar
parameters, the magnetic field, implicitly the rotation (the wind must
be asymptotic already at the surface, which implies slow rotation at the
surface) and the explosion energy, these parameters should be very
nearly the same for all stars.

This corresponds exactly to what has been argued in a very different
context already by G. Bisnovatyi-Kogan in (1970), who based his argument
in turn on a broader suggestion by N. Kardashev (1964),
\cite{BK70,Ka64}.
In this picture, the predecessor star is rotating and has a magnetic
field.  When the core of the star collapses, the core is spun up to a fast
rotating disk, and the collapse along the plane of symmetry stops.  The
potential energy at that stage then is transmitted to the envelope by the
torque of the magnetic fields (i.e. their angular momentum transport), and
that potential energy constitutes the explosion energy.

Therefore, if the data are really well fit by the model, the conclusion is
strongly suggested that not only the concept of magneto-rotational
explosion works, but that in fact it works with very similar numbers in
all stars.

\section{New data}

Since the original detailed quantitative predictions were made, new data
have appeared, which provide very serious tests, e.g. \cite{Hoe02}:

\begin{itemize}
\item{} At present the data set from the KASCADE experiment is the
best across the knee region, and provides a first challenging test.  As we
show below they are very well fit by the data.  Other data also show
consistency, but with large error bars, and with unknown systematics
still, so further conclusions can not be drawn from them.

\item{}  In paper \cite{HirscheggCR} we predicted the energy
dependence of the B/C ratio, from interaction in the thick WR-star
shells, as $E^{-5/9}$.  A refined treatment was given in \cite{CR9}.
As V. Ptuskin then showed some time later, at the ICRC in Salt Lake City,
\cite{Pt99}, the best fit to the data gives an energy dependence of
$E^{-0.54}$, which is quite consistent.

\item{}  In 1997 St. Hunter et al., \cite{Hunter97}, and M. Mori,
\cite{Mori97}, showed that the gamma ray emission from the Galactic
plane could only be fitted by a cosmic ray spectrum, which was
substantially flatter than the one observed.  In \cite{HirscheggCR,CR9} we
suggested that the spectrum could be explained as interaction in the
source region of cosmic rays, specifically the RSG star shells.  The
interaction happens then with a spectrum of $E^{-7/3}$.  We will discuss
the details of this fit elsewhere.

\end{itemize}

\subsection{Specific tests}

The tasks ahead of us are the following, to be quite specific:

\begin{itemize}

\item{}  We normalize all abundances at TeV, using the data collection of
\cite{LB-CR}.

\item{}  We need to remember, that Hydrogen has two components, only one
of which may have a knee (wind-SNe, RSG), the other one has a cutoff
below the knee (ISM-SNe); on the other hand Hydrogen is extremely low in
WR winds.  We need to note here, that the data do not force us at this
time to assume that the knee of the RSG stars and the WR stars is the
same; in fact, it could be, that both knee and cutoff for RSG stars are
quite different, much lower in rigidity than for WR stars. In our first
data fits, described below, we do not yet distinguish RSG and WR stars as
regards the knee and final cutoff.  The KASCADE data are forceful only
for the abundances from the WR stars.

\item{}  Also Helium may have two different observable knees or cutoffs,
corresponding again to RSG supergiants, and WR supergiants, but they also
might be same; we have to keep this uncertainty in mind.

\item{}  We then extrapolate with the predicted spectrum to the knee in
rigidity,  bend at the knee in rigidity, and extrapolate to the cutoff.
As the cutoff is also in rigidity, there is a rolloff of the overall
spectrum, starting from the cutoff in Helium, which we can actually
identify with the ``second knee".

\item{}  We then compare to KASCADE, Akeno, HIRES, ...  One important
point is to remember, that any fit to the data across the knee has to
``reach" the data at higher energy, such as from Akeno, etc.  This
restricts any possible fits quite severely, see \cite{Hoe02}.

\end{itemize}

%%%%%%%%%%%%%%%to insert figure 1 %%%%%%%%
%\begin{figure}
%\centering\rotatebox{90}{\resizebox{16cm}{!}%
%{\includegraphics{:fig-Folder:AV-CReps.eps}}}
%\caption{In this graph we show the fit straight to the elemental curves
%obtained from KASCADE data; A. Vasile; here we use data kindly supplied
%by Zh. Cao from the HiRes collaboration}
%\end{figure}
%%%%%%%%%%%%%%%%%%%%%%%%%%%%%%%%%%%%%%%%%
%

\subsection{Work to be done}

The work is spread among many of our partners, as follows:

\begin{itemize}

\item{}  The detailed abundances are being considered with A. Popescu
(Bukarest), and N. Langer (Utrecht).

\item{}  The energy in explosions is considered with G. Pavalas
(Bukarest).

\item{}  The positron spectra  are done with W. Rhode \etal (Wuppertal).

\item{}  The gamma ray spectrum of the Galaxy is done with S. Casanova
(Bonn), R. Engel (Karlsruhe), W. Rhode (Wuppertal), and many others.

\item{}  The anti-proton spectra are done with E.-S. Seo, R. Sina
(Univ. Maryland), and R. Engel (Karlsruhe).

\item{}  The detailed fit to the KASCADE data is first described in this
paper, in the following main section.

\item{}  Later we will also consider a more detailed fit to the Akeno,
AGASA and HIRES near the ankle with S. Ter-Antonyan (coauthor here), A.
Vasile (coauthor here), Zh. Cao (Utah) and St. Westerhoff (New York).

\end{itemize}

%Samvel's part  %%%%%%%%%%%%%%%%%%%%%%%%%%%%%%%%%%%%%%%%%%%%
\section{Cosmic ray properties from air shower data}

The following section was written by S. Ter-Antonyan.

In general, the relation between energy spectra ($\partial
\Im_{A}/\partial E_{0}$) of
primary nuclei ($A$) and measured EAS size spectra at
observation level ($\Delta
I/ \Delta N_{e,\mu}^*$) is determined by an integral equation

\begin{equation}
\frac{\Delta I(\theta)} {\Delta N_{e,\mu}^*}=
\sum_{A}
\int_{E_{min}}^{\infty}
\frac{\partial \Im_{A}} {\partial E_{0}}
\frac{\partial W(E_{0},A,\theta)}{\partial N_{e,\mu}^*}
dE_{0}
\label{eq:Samvel1}
\end{equation}

\noindent where the kernel function ($\partial W/\partial
N_{e,\mu}^*$) of the equation depends on primary energy,
type of primary nucleus ($A=1,4,\dots,59$), zenith angle
($\theta$), $A-A_{Air}$ interaction model and
response functions of measurements.\\
Eq.~(\ref{eq:Samvel1}) is a typical ill-posed problem and
has an infinite set of solutions for unknown primary energy
spectra. However, the integral Eq.~(\ref{eq:Samvel1}) turns to a
Fredholm equation, if the type of a primary nucleus is defined
directly in the experiment (as it is in the balloon and
satellite measurements \cite{JACEE,RUNJOB} where energy
spectra for different nuclei are obtained up to $10^{15}$ eV).\\
Similarly to \cite{Glass,STPB} we reconstruct
primary energy spectra based on a transformation of the integral
Eq.~(\ref{eq:Samvel1}) to a parametric equation with unknown
spectral parameters. For that, instead of unknown primary
energy spectra of the integral Eq.~(\ref{eq:Samvel1}) we
consider energy spectra according to the predictions
of a multi-component model \cite{CRI,TucsonCR}.
In this case, unknowns are such model parameters as
cut-off and knee energies, spectral indices,
scale factors and fractions of different components.
EAS muon and electron size spectra on the left side of the
Eq.~(\ref{eq:Samvel1}) are taken from recent KASCADE
publications \cite{Ul+01,KAS2}.\\
The kernel function of Eq.~(\ref{eq:Samvel1})
$\partial W/\partial N_{e,\mu}$
is preliminarily calculated by a Monte-Carlo method
provided a given
($A-A_{Air}$) interaction model of primary nuclei and
atmosphere.\\
Solutions of Eq.~(\ref{eq:Samvel1}) for unknown spectral
parameters
at {\emph{a priori}} known primary spectra are easy to
obtain by
means of  $\chi^2$-minimization method \cite{TH,STPB}.
It is clear that solutions found that way are
partial solutions of the integral Eq.~(\ref{eq:Samvel1}).
However, if the values of the spectral parameters are
consistent with the predictions of a primary spectrum model
calculations \cite{CRI}, it will confirm the use
of this model in the investigation energy region.\\
We have already done such investigations \cite{STPB} by
testing a multi-component model of primary cosmic
ray origin with observed EAS size spectra from
KASCADE (1020 g/cm$^2$) and ANI (700 g/cm$^2$) using
experiments in 5 different zenith angular intervals.\\
Here we consider the inverse problem based on KASCADE EAS
electron and "truncated" muon size spectra \cite{Ul+01,KAS2}
in 3 zenith angular intervals. The combined analysis of
electron and muon EAS size spectra allows us to determine the
influence of a nuclear interaction model on the quality of the
inverse problem solution more accurately.

\subsection{Parametrization of the primary energy spectra}

The energy spectra of primary nuclei
according to the multi-component model of primary cosmic ray
origin
\cite{CRI,TucsonCR} at energies $10^{12}-10^{18}$ eV
are presented here in a 3-component form:

\begin{equation}
\frac{\partial \Im} {\partial E_{A}}=
\beta\Phi_{A}\Big(
\delta_{A,1}\frac{d\Im_{1}}{dE_{A}}+
\delta_{A,2}\frac{d\Im_{2}}{dE_{A}}
\Big)
+\Phi_{A}^{EG}\frac{d\Im_{3}}{dE_{A}}
\label{eq:Samvel2}
 \end{equation}

\noindent where the $\beta$ is a dimensionless
normalization parameter,
$\Phi_{A}$ are scale spectral factors
and  model parameters $\delta_{A,i=1,2}$ are
the fractions of each component ($\delta_{A,1}+\delta_{A,2}=1$)
in a primary flux of nuclei (A).\\
The first component (ISM) is
derived from explosions of a normal supernova into an
interstellar medium with expected rigidity-dependent power law
spectra

\begin{equation}
\frac{d\Im_{1}}{dE_{A}} = \left\{
\begin{array}{l@{\quad:\quad}l}
E_{A}^{-\gamma_{1}} &
E_{A} < E_{cut}Z \\
0 & E_A>E_{cut}Z
\end{array} \right.
\label{eq:Samvel3}
\end{equation}

\noindent where the model parameter $E_{cut}Z$ is a cut-off
energy of ISM component at $Z$ nuclear charge.\\
The second component (SW) is a result of explosions of
stars into their former stellar winds with expected
rigidity-dependent power law spectra

\begin{equation}
\frac{d\Im_{2}}{dE_{A}} =
\left\{ \begin{array}{l@{\quad:\quad}l}
E_{A}^{-\gamma_{2}} & E_{A} < E_{k}Z \\
E_{k}^{-\gamma_{2}}(E_{A}/E_{k})^{-\gamma_{3}} & E_A>E_{k}Z \\
0 &  E_{A} > E_{cut}^{SW}Z
\end{array} \right.
\label{eq:Samvel4}
\end{equation}

\noindent where the model parameter $E_{k}Z$ is a knee energy
of SW component and  $E_{cut}^{SW}\simeq2.2\cdot10^5$ TeV is
a corresponding cut-off energy \cite{CRI}.\\
The third, extragalactic (EG) component is approximated also by
rigidity-dependent power law spectra

\begin{equation}
\frac{d\Im_{3}}{dE_{A}} =
\left\{ \begin{array}{l@{\quad:\quad}l}
E_{ank}^{-2.75}(E_{A}/E_{ank})^{-2} &
E_A<E_{ank}Z\\
E_{A}^{-2.75} & E_{A} > E_{ank}Z \\
\end{array} \right.
\label{eq:Samvel5}
\end{equation}

The values of model predictions \cite{CRI} for spectral
parameters are:\\
$\gamma_1=2.75\pm0.04$, $\gamma_2=2.67\pm0.03$,
$\gamma_3=3.07\pm0.1$, $E_{cut}\simeq120$ TV, $E_{k}\simeq 700$
TV, $\Phi_{A=1}^{EG}\simeq0.032$ $(m^2\cdot sec \cdot ster \cdot
TeV)^{-1}$, $E_{ank}\simeq 6.5\cdot 10^5$ TV
at factors of uncertainty $\sim 2$.\\
In \cite{STPB} we have already obtained the evaluations of
spectral indices of ISM and SW components:
$\gamma_1=2.78\pm0.03$, $\gamma_2=2.65\pm0.03$ as solutions
of parametric Eq.~(\ref{eq:Samvel1}) using KASCADE \cite{KAS3} and
\cite{ANI} EAS size spectra at 5 zenith angular intervals.  This is in
quite good agreement with a similar early test \cite{CRIV}. Since these
values agreed with the model predictions here we  set them fixed.

\subsection{EAS size spectra}

In general, the kernel function of Eq.~(\ref{eq:Samvel1}) in EAS
inverse problems is determined by

\begin{equation}
\frac{\partial W}{\partial N_{e,\mu}^*}
\equiv
\int_{\theta _1}^{\theta _2}
\int_{0}^{\infty}
\frac{\partial G(E_{0},A,\theta)}{\partial N_{e,\mu}}
\frac{\partial P(N_{e,\mu})}{\partial N_{e,\mu}^{*}}
\frac{\sin\theta}{\Delta_{\theta}} d\theta dN_{e,\mu}
\label{eq:Samvel6}
\end{equation}

\noindent where $\partial G/ \partial N_{e,\mu}$
is an expected EAS size
electron (truncated muon) spectrum
at a given observation level for $E_{0},A,\theta$
parameters of a primary nucleus and depends on $A-A_{Air}$
interaction model,
$\Delta_{\theta}=\cos\theta_{1}-\cos\theta_{2}$, and
$\partial P/\partial N_{e,\mu}^{*}$ is the error function
of measurements \cite{TH}.\\
We calculated the shower spectra $\partial G/ \partial
N_{e,\mu}$ at KASCADE observation level (1020 g/cm$^2$)
using the CORSIKA6016(NKG) EAS simulation code
\cite{COR} with the QGSJET01 \cite{QGSJET} and SIBYLL2.1
\cite{SIBYLL} interaction models. Input parameters in
simulations were: primary energies $E_A\equiv32,100,\dots
3.2\cdot10^5$ TeV, 4 groups of primary nuclei $A\equiv1,4,16,56$
and 3 zenith angular intervals similarly to KASCADE
experimental data \cite{Ul+01,KAS2}. Intermediate values are
calculated using log-linear interpolations. Relative
statistical errors of calculated shower spectra were
less than $3\%$.\\
The expected detectable EAS size spectra $\partial W/\partial
N_{e,\mu}^*$ were calculated using log-Gaussian approximations
for $\partial G/ \partial N_{e,\mu}$ and error functions
$\partial P/\partial N_{e,\mu}^{*}$ which were obtained
with high accuracy in the investigated energy range.\\
We also investigated the error function of measurements
using CORSIKA code and for reconstructions of
truncated muon sizes we obtained accuracy approximation:
$\Delta N_{\mu}/N_{\mu}\simeq15/\sqrt N_{\mu}+0.05$ that
exceeds the corresponding values of KASCADE approximation
\cite{Glass}. In our calculations we used a more precise
standard option of CORSIKA6016.\\
For the left side of Eq.~(\ref{eq:Samvel1}) we used
KASCADE EAS electron and truncated muon size spectra from
\cite{Ul+01,KAS2} (Fig.~\ref{fig:S-fig2}, symbols). In order
to make a comparison
and subsequent normalization of our data with direct balloon and
satellite measurements at about $10^{14}$ energy range
we also considered the data  from
early publications of KASCADE \cite{KAS3} with corresponding
zenith angular corrections (the first 7 symbols for each angular
interval in Fig.~\ref{fig:S-fig2}a).

\subsection{Results}

Minimization of $\chi^2({\bf{I}}, {\bf{P}})$-functional with
a measurement vector

\begin{equation}
{\bf{I}}\equiv \{\Delta I/\Delta
N_{e,i,k}^*,\Delta I/\Delta N_{\mu,j,k}^*\}
\label{eq:Samvel7}
\end{equation}

\noindent and a corresponding prediction vector ${\bf{P}}$
from the right-hand part of Eq.~(\ref{eq:Samvel1}) was carried out
at $i=1,\dots,42$ energy intervals of EAS electron size spectra,
$j=1,\dots,26$ energy intervals of EAS truncated muon size
spectra and $k=1,2,3$ zenith angular bins from KASCADE data
\cite{Ul+01,KAS2}. However, the combined analysis of electron and
muon size spectra at $\chi^2$-minimization requires to include
in the expected shower spectra
2 additional unknown dimensionless parameters $\eta_e$ and
$\eta_\mu$ which define a constant bias of each
spectrum
due to peculiarities of interaction model and systematic
measurement  errors \cite{TH}.\\
In the first instance, inner spectral parameters
($\gamma_{1,2,3}$, $E_{cut}$, $E_{k}$, $\delta_{A}$, $\Phi_{A}$)
of primary spectra
(\ref{eq:Samvel2}-\ref{eq:Samvel4}) have
been evaluated using normalized
dimensionless EAS size spectra in the measurement vector

\begin{equation}
{\bf{I_n}}
\equiv\Big\{
\frac{1}{U_{e}}
\frac{\Delta I}{\Delta N_{e,i,k}^*},
\frac{1}{U_{\mu}}
\frac{\Delta I}{\Delta N_{m,j,k}^*}\Big\}
\label{eq:Samvel8}
\end{equation}

\noindent where the scalar parameters $U_{e}=\sum_{i,k}(\Delta I/\Delta
N_{e,i,k})$
and $U_{\mu}=\sum_{j,k}(\Delta I/\Delta N_{\mu,j,k})$.
Corresponding
normalizations of
spectra were performed for all components of the prediction vector
${\bf{P_n}}$ at the minimization of the $\chi^2({\bf{I_n}},
{\bf{P_n}})$-functional.
After evaluation of the inner spectral parameters the
values of $\eta_e$ and $\eta_\mu$ were determined  using the
$\chi^2({\bf{I}}, {\bf{P}})$ functional by definition
(\ref{eq:Samvel6}) at fixed inner spectral parameters.\\
Finally, the dimensionless parameter $\beta$ in expression
(\ref{eq:Samvel2}) was  determined by a normalization of the
all-particle spectrum obtained
$\sum (\partial\Im/\partial E_A)$ with
JACEE data at 100 TeV energy.

\begin{table}
\begin{center}
\begin{tabular}{ccccc}
\hline
Spectral      &   QGSJET01  & SIBYLL2.1
&3-component  &Comments \\
Parameters    &             &
&predictions  &  \\
\hline
$\gamma_1$    &2.78$\pm$0.03&     -
&2.75$\pm$0.04&\cite{STPB}\\
$\gamma_2$    &2.65$\pm$0.03&     -
&2.67$\pm$0.03&\cite{STPB}\\
$\gamma_3$    &3.25$\pm$0.04&3.25$\pm$0.04&3.07$\pm$0.1
&3.28$\pm$0.07\cite{STPB}\\
$E_{cut}$         &200$\pm$100  &200$\pm$100  &120-250
&210$\pm$60 $TV$ \cite{STPB}\\
$E_k$         &2100$\pm$140 &1910$\pm$150 &700-1400
&1900$\pm$100 $TV$ \cite{STPB}\\
$\delta_{A=1,2}$&0.47$\pm$0.04&0.5$\pm$0.04 &   -
&0.5-0.8 \cite{STPB}\\
$\delta_{A>1,2}$&     0.9     &    0.9      &    -
&0.85-1 \cite{STPB}\\
$\beta\Phi_P$      &0.120$\pm$0.007&0.106$\pm$0.006&
&$(m^2\cdot s\cdot sr\cdot TeV)^{-1}$\\
$\beta\Phi_{He}$   &0.089$\pm$0.011&0.084$\pm$0.010&
&$(m^2\cdot s\cdot sr\cdot TeV)^{-1}$\\
$\beta\Phi_{O}$    &0.058$\pm$0.007&0.064$\pm$0.006&
&$(m^2\cdot s\cdot sr\cdot TeV)^{-1}$\\
$\beta\Phi_{Fe}$   &0.026$\pm$0.005&0.035$\pm$0.005&
&$(m^2\cdot s\cdot sr\cdot TeV)^{-1}$\\
\hline
\end{tabular}
\caption{Spectral parameters of reconstructed energy spectra
(\ref{eq:Samvel2}-\ref{eq:Samvel4})  for 4 groups of primary
nuclei in the framework of a 3-component model of primary
cosmic ray origin and QGSJET and SIBYLL interaction models
based on KASCADE date.}
\end{center}
\end{table}

\begin{figure}
\centering\rotatebox{0}{\resizebox{15cm}{!}%
{\includegraphics{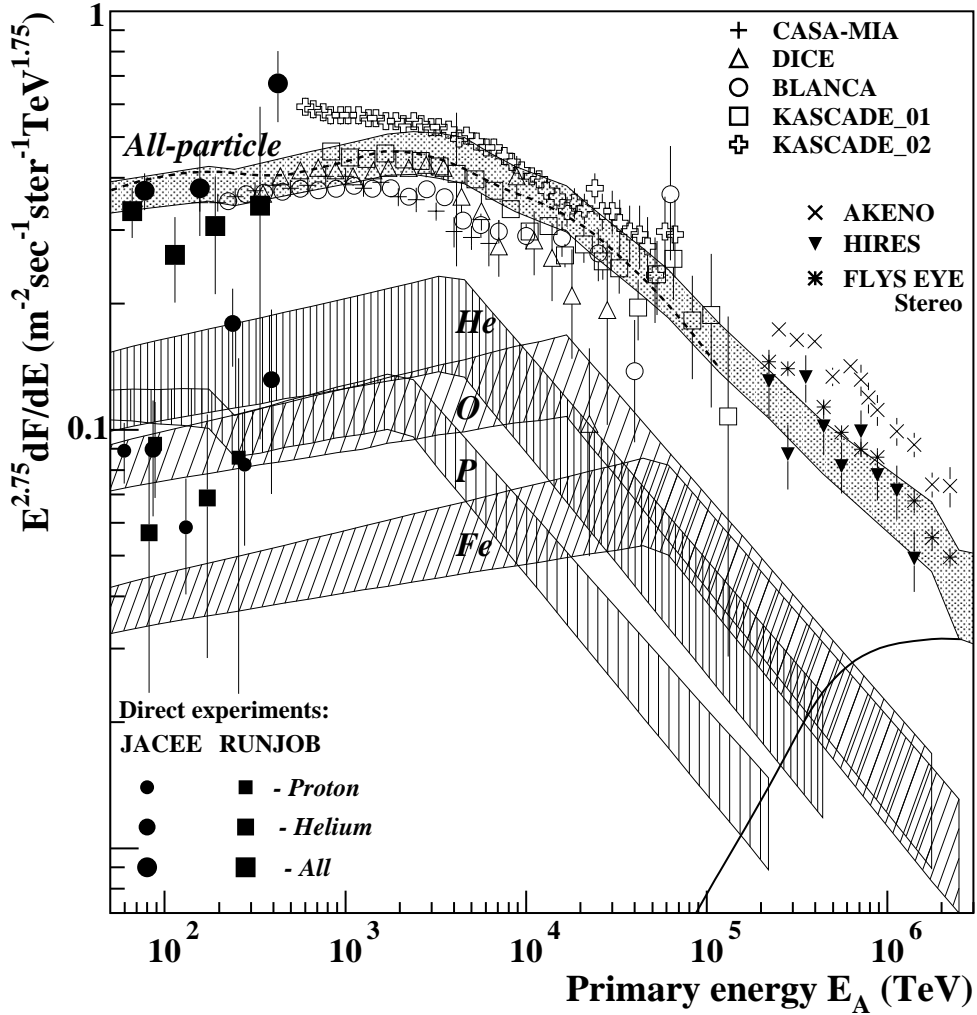}}} \caption{\label{fig:S-fig1}
Expected all-particle energy spectrum and energy spectra of
different nuclei (shaded areas) according to a 3-component model
of primary cosmic rays. The solid line is the expected (third)
extragalactic component and the dashed lines correspond to the
all-particle spectrum from \cite{STPB}. The JACEE, RUNJOB, CASA,
DICE, BLANCA, KASCADE02 data is taken from the review
\cite{Swordy},  KASCADE01 - from \cite{Ul+01}.}
\end{figure}

Table 1 contains the values of all spectral parameters
which were obtained by the method above
at QGSJET01 \cite{QGSJET} and SIBYLL2.1 \cite{SIBYLL}
interaction models and KASCADE EAS data \cite{Ul+01,KAS2}.
The normalization  factors in Table 1 are:
$\beta=1.13\pm0.05$ at QGSJET model and
$\beta=1.0\pm0.05$ SIBYLL model.\\
The energy spectra obtained of different nuclei and the
corresponding all-particle spectrum in comparison with
JACEE  \cite{JACEE} and RUNJOB \cite{RUNJOB} direct
measurements and  EAS data \cite{Swordy} up to energy
$3\cdot10^{18}$ eV are presented in  Fig.~\ref{fig:S-fig1}.

\begin{figure}
\centering\rotatebox{0}{\resizebox{15cm}{!}%
{\includegraphics{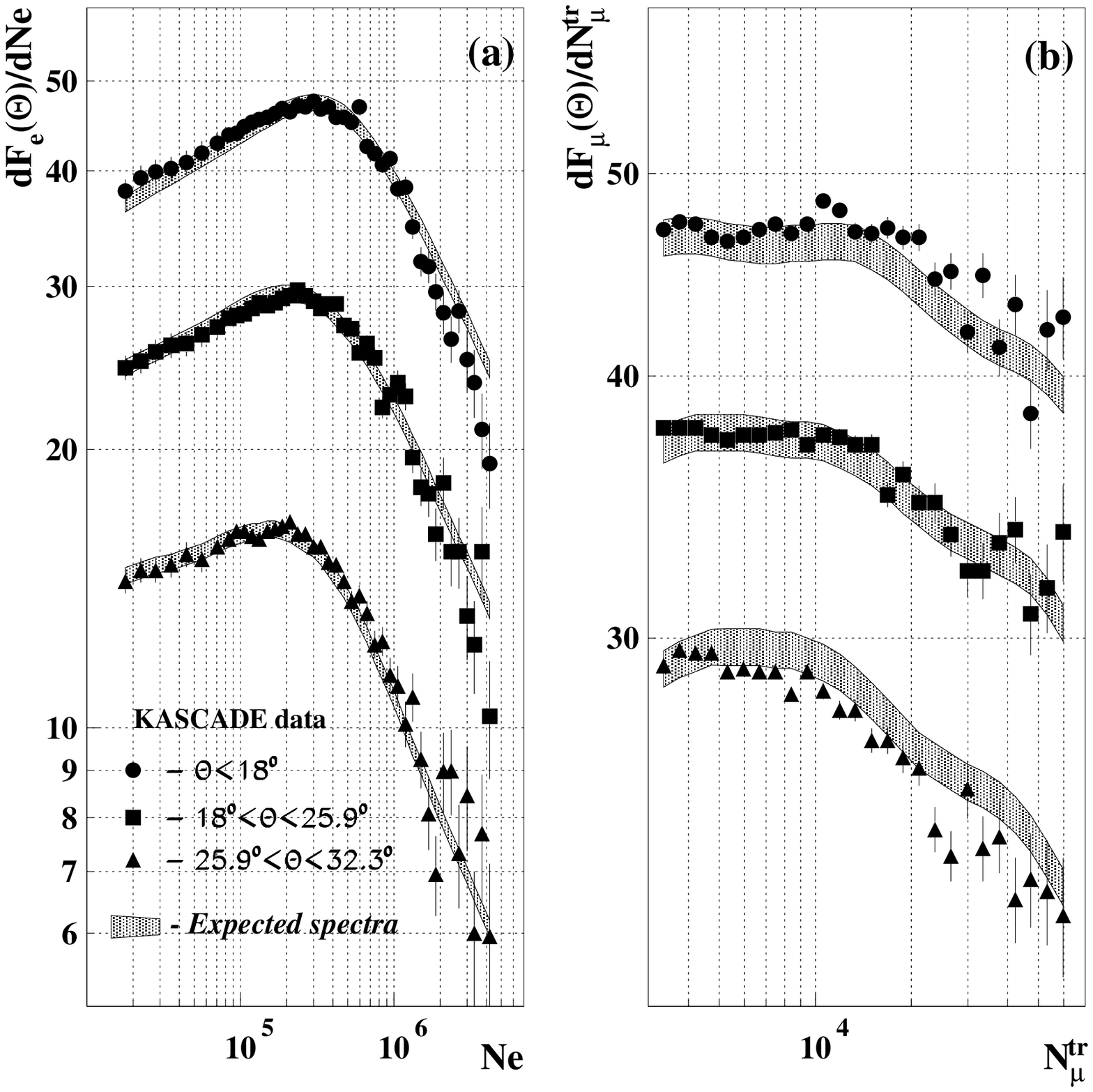}}} \caption{\label{fig:S-fig2}
KASCADE EAS electron (a) and truncated muon (b) size spectra
(symbols) at 3 zenith angular bins (from \cite{Ul+01}) and
corresponding expected spectra (shaded areas) in the framework of
a 3-component model of primary cosmic ray origin and the QGSJET
interaction model.}
\end{figure}

The thick line corresponds to the expected energy
spectra of the extragalactic component
\cite{CRI,CRIV,UHECRI,UHECRII}. This  flux is also included
in the all-particle spectrum in Fig.~\ref{fig:S-fig1}\\
The expected EAS electron and truncated muon size spectra
corresponding to our primary spectra obtained with the QGSJET
interaction model in comparison with the KASCADE data
\cite{Ul+01,KAS2} are shown in Fig.~\ref{fig:S-fig2}a,b.

\begin{figure}
\centering\rotatebox{0}{\resizebox{15cm}{!}%
{\includegraphics{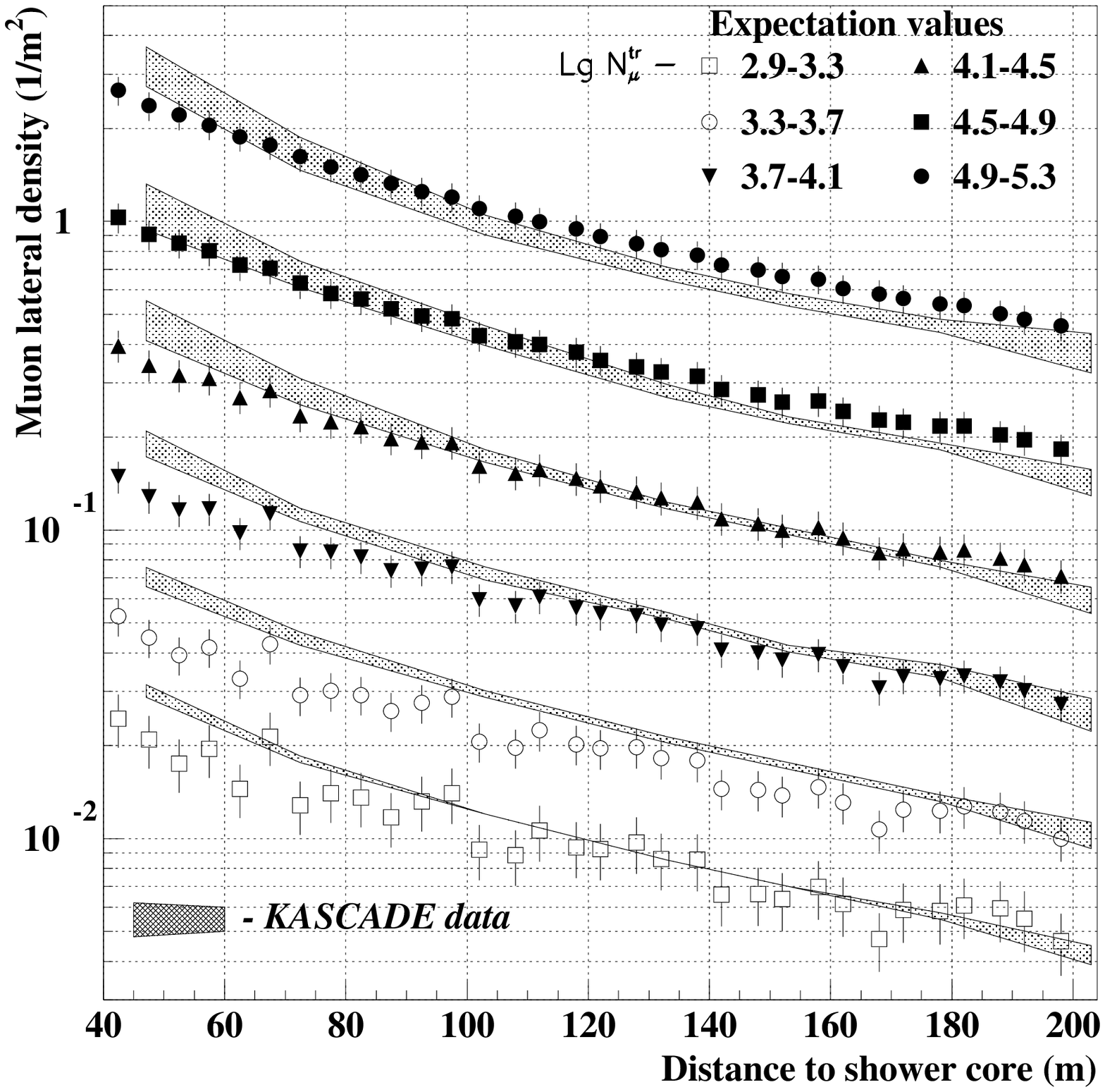}}} \caption{\label{fig:S-fig3}
KASCADE muon lateral distribution functions (shaded areas) at
different truncated muon sizes and the corresponding 3-component
model predictions with the QGSJET interaction model (symbols).}
\end{figure}

The value of  $\chi^2\simeq1$ with 2$\%$ uncertainty (the
width of a shaded area) of expected data
(and 4$\%$ uncertainty at SIBYLL model).
For biases  of shower
spectra (Fig.~\ref{fig:S-fig2}a,b) we obtained the
values: $\eta_e=1.23\pm0.04$,
$\eta_{\mu}=1.12$ at QGSJET model and $\eta_e=1.0\pm0.01$,
$\eta_{\mu}= 1.3\pm0.02$ at SIBYLL model. These values point
out the possible existence of systematic biases both in the EAS
measurements and in the interaction models.\\
In  Fig.~\ref{fig:S-fig3} the expected muon lateral distribution
functions  (symbols) and corresponding KASCADE data
shaded area) from  \cite{KAS4} at different intervals of
detected truncated muon size are shown.

\begin{figure}
\centering\rotatebox{0}{\resizebox{15cm}{!}%
{\includegraphics{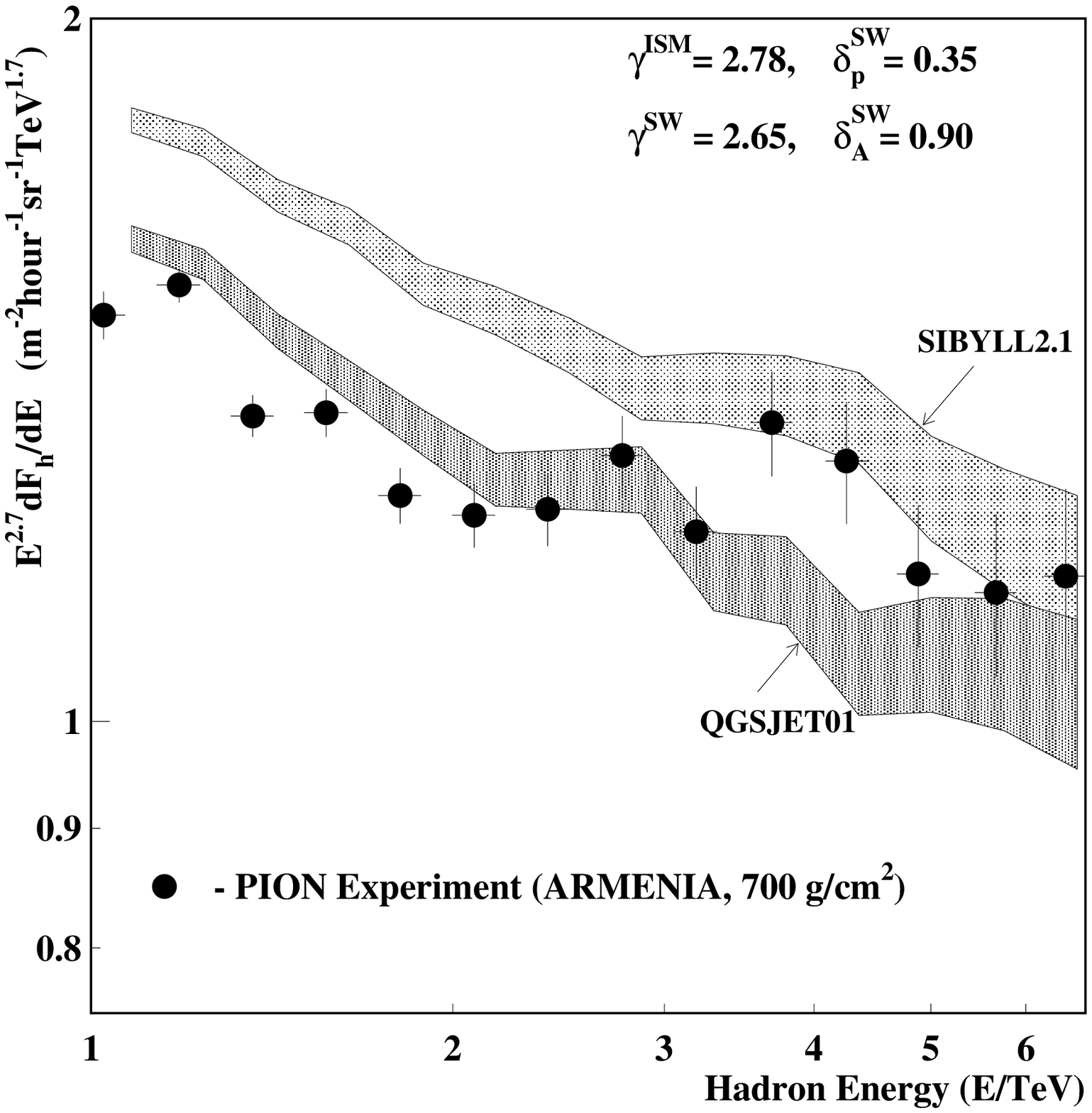}}} \caption{\label{fig:S-fig4}
Vertical hadron energy spectrum at mountain level (700 g/cm$^2$)
obtained by the PION experiment (symbols) and the expected spectra
(a shaded  areas) in the framework of the 3-component primary
model and QGSJET and SIBYLL interaction models.}
\end{figure}

In order to test the behavior of primary energy spectra obtained
in a larger energy range we also calculated the expected hadron
energy spectra at mountain level for comparison with
data of the PION experiment (3200m a.s.l.) \cite{PION,Samo}
at 1-7 TeV hadron energies at observation level (700 g/cm$^2$).
The calculation was carried out using the QGSJET and SIBYLL interaction
models and using the method introduced above without normalizations
($\beta=1$).  The best agreement of the expected hadron
flux at mountain level in the framework of the 3-component primary
model predictions and PION data was achieved at
$\delta_{A=1,2}=0.35\pm0.05$ and $\delta_{A>1,2}=0.9\pm0.1$
fractions of component. These results are shown in Fig.~\ref{fig:S-fig4}.
The effective primary energy range corresponding to these data is
$E_0\sim10-100$ TeV/nucleon.\\

Thus, predictions of the multi-component model of the cosmic ray
origin \cite{CRI,TucsonCR} explain the measured KASCADE EAS data in the
knee region ($E_0\simeq10^{14}-10^{17}$ eV) and hadron spectra at
mountain level ($E_0\simeq10^{13}-10^{14}$ eV) with an accuracy of
10-15$\%$ in the framework of the QGSJET and SIBYLL interaction  models
respectively.\\ The rigidity-dependent behavior of spectra
for different primary nuclei is the same for the two interaction
models.\\
The agreement of the expected all particle spectrum and world
data in
$E_0\simeq10^{17}-10^{18}$ eV primary energy range displays the
presence
of the extragalactic component of primary cosmic rays in accordance with
the 3-component model predictions.
%End of Samvel's part %%%%%%%%%%%%%%%%%%%%%%%%%%%%%%%%%%%%%%%%%%%%

\section{The fit and consequences}

Here we have described the actual fit to the data; we did this in two
steps, first just a direct fit to the KASCADE elemental data as
published, as done by A. Vasile, and then by an extensive Monte Carlo run
to consider all published detailed distributions from KASCADE directly,
described in the section above written by S. Ter-Antonyan.  We refer
here to the ICRC publications, such as \cite{Ul+01,Hoe+01}.  An
independent, but similar approach is described by G. Schatz,
\cite{Schatz02}; recent KASCADE papers are
\cite{KAS5,KAS6,Hoe02,KAS7}.  It is gratifying, that a first direct fit,
as done by A. Vasile, and the extensive fitting done by S. Ter-Antonyan,
not only agree with each other, but actually agree with the prediction,
and with the early tests,
\cite{CRIV}.

We summarize, that the KASCADE data as well as a first approximation to
the HiRes, Akeno and Yakutsk data can be fully fitted by the model with
parameters very close to what had been predicted.  We need to
emphasize here, that we checked the consistency of the predicted
model with the data; we did not check the uniqueness of the solution
found.  We do note that no other model has been either inferred nor
proposed, which fits all the same data, but it is quite conceivable
that there is such a model, yet to be worked out or published.  The
approaches by M. Malkov, J.-P. Meyer and L. O'C. Drury, E. Berezho et al.,
and I. Axford et al. all provide a different approach to the quest for the
origin of cosmic rays.  The various approaches may yet converge to a
common theory.

Focussing now on the concept originally started some time ago, there are a
number of important conseqences from the success of the model proposed:

\subsection{A common final state for massive stars}

The are several very interesting consequences from the success of this
approach.  Some of these consequences follow by necessity, which entails
if these consequences can be falsified, then the original model
fails.  We will identify these critical steps below.

\begin{itemize}

\item{} The abundances in stellar winds near to and beyond 20 solar
masses, RSG and WR supergiants dominate the cosmic ray chemical
composition -- this is in contrast to the superbubble concept; however,
since the superbubbles in fact also have a mixture of wind abundances,
these two approaches may be consistent in this point with each other.  The
argument, that the abundances of cosmic rays derive from accelerated dust
particles is starkly different, \cite{JPMeyer97}.

\item{} The energy in the cosmic ray population of Helium, Carbon, and
Oxygen is so large as to require a much larger energy content in
cosmic rays per supernova; note that the RSG and WR stars are the dominant
contributors, stars which are quite rare.  The energy requirement in
cosmic rays per supernova of this kind is then about $10^{51}$ ergs.

\item{} If we allow for an overall efficiency of 10 \% of putting
explosion energy into cosmic rays, the explosion energy is then
required to  be $10^{52}$ ergs for such supernovae.

\item{} The most important conclusion from the approach outlined here
is that stars near to or above 25 solar masses explode due to the
magneto-rotational mechanism.  We cannot say for sure at this stage at
which exact mass we have a transition in supernova mechanisms to the
magneto-rotational mechanism, since the cosmic ray argument is really
clear only for the WR stars.

\item{}  Knee energy and cutoff energy need to be the same for
all supernovae contributing.  This is easiest if those stars form
a ``common final state", as regards rotation, magnetic fields, mass,
explosive energy.

\item{} There is first evidence for an explosion energy of order $10^{52}$
ergs from a very massive star from SN1998bw, \cite{Ho99,Na01}.
That specific supernova was also highly asymmetric, with a rotational
symmetry quite possible, consistent with the mechanism proposed by
Bisnovatyi-Kogan.

\end{itemize}

\subsection{The magneto-rotational mechanism}

The magnetorotational supernova mechanism was suggested by
Bisnovatyi-Kogan, \cite{BK70}.  The idea was to get the energy for the
explosion from the rotational (gravitational) energy using the magnetic
field. Most stars are rotating differentially. The magnetic field is
``frozen" into the ionized gas of the star. Then differential rotation
leads to an amplification of the toroidal component of the magnetic
field. When magnetic pressure becomes comparable with the gas pressure
a compression wave forms and moves out to the envelope of the star.
When moving along a steeply decreasing density profile this wave
transforms to a strong fast MHD shock which pushes part of the envelope
of the star to the outside, and so forms an explosion. Simulations of the
magnetorotational mechanism for a magnetized cloud \cite{ABKM00,MABK03}
show that this mechanism is rather effective.  The simulations of the
collapse of a rotating pre-supernova star show that after the short
collapse stage the star consists of a rapidly rotating core and a slowly
rotating envelope.  At the transition region between the core and the
envelope the rotation is very differential and in such a situation the
inclusion of even a weak initial magnetic field could lead to
amplification of the toroidal component and so to a magnetorotational
explosion.  Further work is in \cite{BKM92,ABK96,ABKM01}.

\subsection{Implications for Gamma Ray Bursts?}

There is an interesting resemblance of this explosion mechanism to the
concept of a hypernova developed in the context of Gamma Ray Bursts,
\cite{Pa98,Piran99}.  Using the concept, \cite{GRB1,GRB2}, does indeed
suggest an energy scale of also $10^{52}$ erg, derived in that specific
model as an upper energy limit implied by a fit to the fluence
distribution, allowing for observational selection effects.  The recent
observations by Schaefer \etal, \cite{Sch02}, support the concept that at
least some GRBs blow up into a stellar wind.

This leads then one more time to the question of what the relation is
between Gamma Ray Burst explosions and Supernova explosions, a question
explored by many.

There is much recent work also by C. Wheeler, \etal,
\cite{AWM02,Ak02,Ho02,Wa02,Wh02}.

\subsection{A bright standard candle for cosmology?}

There is a corollary also for cosmology:  If the explosion energy is the
same for all supernovae above about a zero age main sequence mass of
near
25 solar masses, as suggested by the arguments above, one may speculate
even further:  Could it be that also the maximum luminosity, or the
integral over the luminosity time curve is the same for all such
supernovae?  There is the obvious problem, that these explosions are
rotationally symmetric only, and so present vastly different views to
observers at different angles to the symmetry axis.  This may be resolved
by infrared observations, and also by polarization observations, just as
for active galactic nuclei, for which infrared observations give the best
approximation for an isotropic emission, and polarization observation do
give clues for the angle between symmetry axis and line of sight.

If we could thus derive from observations the maximum luminosity or its
time integral, we might have a much brighter standard candle for use in
cosmology than the supernovae of type Ia, which derive from the collapse
of a white dwarf in a stellar binary system, and have a fairly low
explosion energy.

\subsection{Transition: Beyond GZK cutoff}

Let us finish with a brief note on the events beyond the Galaxy, the
events with an energy beyond $3 \, 10^{18}$ eV.  Recent reviews are in,
e.g., \cite{B97,Bhatta99}.  The latest data and their discussions are in
\cite{HiRes02a,HiRes02b} for HiRes, and in
\cite{Ta+98,Ta+99,Hay+00,Ide+01,Yo+02a,Yo+02b} for AGASA.

The main dispute at the present time is the apparent discrepancy between
the AGASA data and the HiRes data, with the HiRes data suggesting a
downturn of the overall spectrum in qualitative agreement with the
GZK-cutoff concept, although at a somewhat higher energy, as would be
expected from a highly inhomogeneous source distribution, such as any
subpopulation of galaxies.  On the other hand, the AGASA data suggest a
continuation of the spectrum to higher energies, which would require a
new component to appear from under the lower flux contributors, in
perfect agreement with a scenario involving the decay of topological
defects.  The real discrepancy between the two data sets is only about 2
sigma, as discussed at the meeting in Paris June 2002.  It is widely
expected that AUGER data will resolve this issue, and then maybe the
further experiments will be required, such as EUSO and OWL, the European
and American space missions to detect airshowers at extremely high
energies.

\section{Conclusions}

The successful fit of the new KASCADE data, under the constraint to also
fit the higher energy data from HiRes, AGASA and Yakutsk, with the cosmic
ray source model originally proposed some time ago, and developed further
since then, leads to a number of important conclusions. The most important
points are:

\begin{itemize}

\item{}  The origin of Galactic cosmic rays may be much closer to a full
understanding of their origin.  Sofar all quantitative tests, which have
been made, show consistency with the model proposed earlier.  But much
work remains to be done.

\item{}  All stars above a zero age main sequence mass of about 25 solar
masses converge to a ``common final state".

\item{}  Those stars explode with about $10^{52}$ ergs in a highly
anisotropic explosion, following the mechanism originally proposed by
G. Bisnovatyi-Kogan more than 30 years ago, which was then based on a
broader suggestion by Kardashev, involving rotation, magnetic fields and
potential energy.

\item{}  These supernovae may constitute a new very bright standard
candle, useful in cosmology, provided we could determine their luminosity
integrated over $4 \pi$ from infrared, polarization or other observations.

\end{itemize}

\section{Acknowledgements}

P.L. Biermann would like to acknowledge the hospitality, first at the
University of Maryland, offered by his colleague Eun-Suk Seo in the
early spring of 2002, and then during his prolonged stay at the
University of Paris VII, in the late spring 2002, offered by  his
colleagues Norma Sanchez and Hector de Vega, as well as their hospitality
at many meetings in Erice, Paris and Palermo.  PLB would especially like
to acknowledge a day-long debate with I. Axford at the Taiwan meeting in
April 2002, arguing about the agreements and also differences between the
superbubble concept and the stellar wind concept. PLB would also like to
express his appreciation to N. Langer for a similarly long discussion at
Utrecht in May 2002, arguing about supernova physics.  And, PLB would
also like to acknowledge many discussions with G. Schatz on the
intricacies of trying to elucidate cosmic ray spectra from shower data.
The material in this paper was the topic of a long helpful debate with T.
Stanev in Paris in the spring of 2002, for which PLB is also very
grateful.   Help with some of the data was generously supplied by Zh. Cao
from Utah and D. Bergman from Rutgers.  The authors would like to
acknowledge with gratitude discussions with many friends and colleagues,
especially E.J. Ahn, I. Axford, Zh. Cao, S. Casanova, M. Chirvasa, A.
Donea, R. Engel, T. En{\ss}lin, H. Falcke, C. Galea, R. Engel, A. Haungs,
H. Kang, M. Kaufman, P.P. Kronberg, N. Langer, A. Lazarian, H. Lee, K.
Mannheim, S. Markoff, G. Medina-Tanco, A. Meli, F. Munyaneza, A. Olinto,
B. Nath, G. Pavalas, A. Popescu, R. Protheroe, G. Pugliese, J. Rachen, W.
Rhode, G. Romero, E. Roulet, D. Ryu, N. Sanchez, K. Sato, G. Sch{\"a}fer,
G. Schatz, H. Seemann, E.-S. Seo, R. Sina, T. Stanev, V. Tudose, H. de
Vega, Y.P. Wang, A. Watson, T. Weiler, St. Westerhoff, and Ch. Zier.
Work with PLB is being supported through the AUGER theory and membership
grant 05 CU1ERA/3 through DESY/BMBF (Germany); further support for the
work with PLB comes from the DFG, DAAD, Humboldt Foundation (all
Germany), grant 2000/06695-0 from FAPESP (Brasil) through G.
Medina-Tanco, a grant from KOSEF (Korea) through H. Kang and D. Ryu, a
grant from ARC (Australia) through R.J. Protheroe, and European INTAS/
Erasmus/ Sokrates/ Phare grants with partners V. Berezinsky, M.V. Rusu,
and V. Ureche.  All these sources of support are gratefully
acknowledged.  The work reported here was also specifically supported by
the DAAD through a fellowship to S. Ter-Antonyan.


\begin{thebibliography}{999}

\bibitem{HiRes02a} Abu-Zayyad, , \etal, (2002), astro-ph/0208243
%  The HiRes results
%

\bibitem{AWM02}  Akiyama, Sh., Wheeler, J.C., Meier, D.L.,
Lichtenstadt, I., \ApJ  {\bf  } (in press) (2002), astro-ph/0208128
%  Title: The Magnetorotational Instability in Core Collapse Supernova
%  Explosions
%

\bibitem{Ak02}  Akiyama, Sh., Wheeler,J.C., in {\it proc. "Core Collapse
of Massive Stars}, ed. C. L. Fryer, Kluwer Academic Publ., (2002),
astro-ph/0211458
% Title: Magnetic Field in Supernovae
%

\bibitem{KAS4}  Antoni, T., \etal, {\it Astropart.Phys.} {\bf{14}},  245
(2001)
%

\bibitem{KAS5}  Antoni, T., \etal, {\it Astropart.Phys.} {\bf{16}},
373-386 (2002)
%   Title = Muon density measurements with the KASCADE central
%  detector
%

\bibitem{KAS6}  Antoni, T., \etal, {\it Astropart.Phys.} {\bf{16}},
245-263 (2002)
%   Title = A non-parametric approach to infer the energy spectrum and
%  the mass composition of cosmic rays
%

\bibitem{KAS7}  Antoni, T., \etal, {\it Astropart.Phys.} {\bf{18}},
319-331 (2003)
%   Title = The information from muon arrival time distributions of
%  high-energy EAS as measured with the KASCADE detector
%

\bibitem{RUNJOB} Apananseko, A.V., \etal, {\it Proc. 26th ICRC, Salt
Lake City} {\bf{3}},  163 (1999)
%

\bibitem{ABK96}  Ardeljan, N.V., Bisnovatyi-Kogan, G.S., Kosmachevskii,
K.V., Moiseenko, S.G., {\it Astron. \& Astrophys. ASuppl.}  {\bf  115 },
573 (1996)
%  Title:  An implicit Lagrangian code for the treatment of nonstationary
%  problems in rotating astrophysical bodies.,
%

\bibitem{ABKM00}  Ardeljan, N.V., Bisnovatyi-Kogan,
G.S., Moiseenko, S.G., \AA {\bf  355}, 1181 - 1190 (2000)
%  Title:   Nonstationary magnetorotational processes in a rotating
%  magnetized cloud.
%

\bibitem{ABKM01}  Ardeljan, N.V., Bisnovatyi-Kogan, G.S., Moiseenko, S.G.,
in {\it Proc. of XX Texas Symposium of Relativistic astrophysics.
Austin,AIP Conf. Proc.} {\bf 586}, 433 (2001)
%  Title:  Magnetorotational explosion. Results of 2D simulations.
%

\bibitem{JACEE} Asakimory, K. \etal, \ApJ  {\bf 502}, 278 (1998)
%

\bibitem{Ax94}  Axford, W. I., \ApJ {\it Suppl.}  {\bf 90}, 937 - 944
(1994)
%  Title:  The origins of high-energy cosmic rays
%  talks about knee and beyond
%

\bibitem{PION}  Avakyan, V.V., \etal,  {\it Yadernaya Fizika}
{\bf{50}},  134 (1989), {\it Soviet Journal of Nuclear Physics}, in
Russian)
%

\bibitem{BZ34}  Baade, W., Zwicky, F.,  1934,  {\it Proc.
    Nat. Acad. Science}, {\bf 20}, no. 5, 259 - 263.
%

\bibitem{Bell78}  Bell, A.R., 1978 \MNRAS  {\bf 182}, 147 - 156, and 443
- 455.
%

\bibitem{Venyabook} Berezinskii, V.S., {\it et al.},
``Astrophysics of Cosmic Rays",  North-Holland, Amsterdam (especially
chapter IV) (1990).
%

\bibitem{Bhatta99}  Bhattacharjee, P. \& Sigl, G.,
   {\it Physics Reports},  {\bf 327}, 109 - 247 (2000),
    astro-ph/9811011.
% general review of UHECRs and astroparticle physics
%

\bibitem{B97} Biermann, P.L., 1997,  {\it Journal of Physics}
{\bf G 23}, 1.
%  UHECR review

\bibitem%[Biermann 1993]
{CRI} Biermann, P.L., \AA {\bf 271}, 649 (1993), astro-ph/9301008.
%  Title:  Cosmic rays I. The cosmic ray spectrum between
%  $10^4$ GeV and $3\, 10^9$ GeV
% in STA-list as {PeBi1} ################

\bibitem%[Biermann \& and Cassinelli 1993]
{CRII}  Biermann, P.L. \& and Cassinelli, J.P.,
\AA {\bf 277}, 691 (1993), astro-ph/9305003.
%  Cosmic rays II. Evidence for a magnetic rotator Wolf-Rayet
%    star origin
%

\bibitem{CRIII} Biermann, P.L.,  \& Strom, R.G., \AA {\bf 275}, 659
(1993), astro-ph/9303013
%  Cosmic Rays III. The cosmic ray spectrum between 1 GeV and
%  $10^4$ GeV and the radio emission from supernova remnants,

\bibitem{CalgaryCR}  Biermann, P.L.,  invited plenary lecture at the
23rd International Conference on Cosmic Rays, in Proc. ``Invited,
    Rapporteur and Highlight papers"; Eds. D.
    A. Leahy et al., World Scientific, Singapore, 1994, p. 45
%  Title:  Cosmic rays: origin and acceleration - what can we
%  learn from radio astronomy,
%

\bibitem{TucsonCR} Biermann, P.L., in {\it Cosmic winds and the
Heliosphere}, Eds. J. R. Jokipii et al., Univ. of Arizona press, p. 887 -
957 (1997), astro-ph/9501030.
%  Title: Supernova blast waves and pre-supernova winds: Their cosmic
%  ray contribution
%

\bibitem{VulcanoCRa}  Biermann, P.L., review at the Vulcano
meeting on ``Frontier objects in astrophysics and particle physics",
Eds. Giovannelli et al., Italian Physical Society, Conf. Proc.
{\bf 47}, Bologna 1995, p. 469
%  Title:  Production and acceleration of cosmic rays,
%

\bibitem{VulcanoCRb} Biermann, P.L., some concluding remarks at
the Vulcano meeting on Frontier objects in astrophysics and particle
    physics, Eds. Giovannelli et a., Italian Physical Society, Conf.
    Proc. {\bf 47}, Bologna 1995, p. 593
%  Title:  On the knee of the cosmic ray spectrum,
%

\bibitem{HirscheggCR}  Biermann, P.L., invited lecture at the Nuclear
Astrophysics meeting at Hirschegg, 1997, in Proc., GSI, Darmstadt, p. 211
- 222, 1998
%  Title:  Cosmic ray interactions in the Galaxy
%

\bibitem{CR9}  Biermann, P.L., Langer, N., Seo, E.-S., Stanev, T., \AA
{\bf 369}, 269 - 277 (2001)
%  Title:  Cosmic Rays IX. Interactions and transport of
%   cosmic rays in the Galaxy,
%

\bibitem{BK70}  Bisnovatyi-Kogan, G.S., {\it Astron.Zh. (Sov.
Astron.)},  {\bf 47}, 813 (1970)
%

\bibitem{BKM92}  Bisnovatyi-Kogan, G.S., Moiseenko, S.G., {\it Sov.
Astron.}  {\bf  36} 285, (1992)
%  Title:  Violation of mirror symmetry of the magnetic field in a
%  rotating star and possible astrophysical manifestations.,
%

\bibitem{BO78}  Blandford, R. D.; Ostriker, J. P., \ApJL  {\bf  221},
L29 - L32 (1978)
%  Title:  Particle acceleration by astrophysical shocks
%

\bibitem{ANI}  Chilingaryan, A.A., \etal, {\it Proc. 26th ICRC, (Salt
Lake  City) }  {\bf{1}},  240 (1999)
%

%\bibitem{Clay98} Clay, R. \& Dawson, B., 1998, ``Cosmic Bullets"
%  (paperback), by Allen \& Unwin (Australien), preceded by a cloth-bound
%  edition.
%

\bibitem{Drury83}  Drury, L. O'C., {\it Rep. Progr. Phys.}  {\bf
46}, 973 - 1027 (1983)
% Title:  "An introduction to the theory of diffusive shock acceleration
%  of energetic particles in tenuous plasmas"

\bibitem{Fermi49}  Fermi, E.,    {\it Phys. Rev. }
    2nd ser., {\bf 75}, no. 8, 1169 - 1174 (1949)
%

\bibitem{Fermi54}  Fermi, E.,  {\it Astrophys.J.}
{\bf 119}, 1 - 6 (1954)
%

\bibitem{SIBYLL} Fletcher, R.S.~, Gaisser, T.K.~, Lipari, P.~, \&
Stanev, T.~,  {\it Phys.Rev. D} {\bf{50}},  5710 (1994) / Engel, J.,
Gaisser, T.K.~, Lipari,  P.~, Stanev, T.~, {\it Phys.Rev.D.} {\bf{46}},
5013 (1992) / Engel, R., \etal, {\it Proc. 26th ICRS (Salt Lake City)},
{\bf{1}}, 415 (1999)
%

\bibitem{Gaisser90}  Gaisser, T.K.,   {\it Cosmic Rays and Particle
    Physics}, Cambridge Univ. Press (1990)
%

\bibitem{GM77}  Garcia-Munoz, M., Mason, G. M., Simpson, J. A., \ApJ
{\bf  217},  859 - 877 (1977)
%  Title:  The age of the galactic cosmic rays derived from the abundance
%  of Be-10
%

\bibitem{GM87}  Garcia-Munoz, M., \etal, \ApJS  {\bf  64}, 269 - 304
(1987)
%  Title;  Cosmic-ray propagation in the Galaxy and in the heliosphere -
%  The path-length distribution at low energy
%

\bibitem{GS63}  Ginzburg, V.L. \& Syrovatskii, S.I.,
 {\it The origin of cosmic rays}, Pergamon Press, Oxford (1964),
 Russian edition (1963).
%

\bibitem{Glass} Glasstetter, R., \etal, {\it Proc. 26th ICRC Salt
Lake City} {\bf{1}} 222 (1999)
%

\bibitem{KAS3}  Glasstetter, R., \etal,  {\it Nucl.Phys. B (Proc.Suppl)}
{\bf  75A}, 238 (1999)
%

%\bibitem{HMR00}  Harari, D., Mollerach, S., Roulet, E., \JHEP
%{\bf 0010}, 047 (2000), astro-ph/0005483
%  Title:  MAGNETIC LENSING OF EXTREMELY HIGH-ENERGY COSMIC RAYS
%  IN A GALACTIC WIND.
%

\bibitem{KAS2}  Haungs, A. {\it 18th ECRS}, Moscow (2002) Ep1a.2,/ \\
http://dbserv.npi.msu.su/~18sym/conference/transparencies.phtml/\\
K.-H.~Kampert et al. (KASCADE collaboration),
Highlight Paper, {\it Proc. 27th ICRC, Hamburg} (2001), astro-ph/0204205
(2002)

\bibitem{Hay+00}  Hayashida, N., \etal, (2000), astro-ph/0008102,
appendix for Takeda \etal., \ApJ {\bf 522}, 225 - 237 (1999)
%  Title:  Updated AGASA event list above 4x10^19eV
%

\bibitem{COR}  Heck, D., Knapp, J., Capdevielle, J.N., Schatz,
G., \& Thouw, T., {\it Forschungszentrum Karlsruhe Report}, FZKA Nr. 6019
90 p. (1998)


\bibitem{Hillas84}  Hillas, A.~M., {\em Ann. Rev. Astron. Astrophys.}
{\bf 22}, 425 (1984).
%

\bibitem{HiRes02b} HiRes-Coll., \ApP (submitted) (2002), astro-ph/0208301
%  Title: Measurement of the Spectrum of UHE Cosmic Rays by the FADC
%  Detector of the HiRes Experiment
%

\bibitem{Ho99}  H{\"o}flich, P., Wheeler, J. C., Wang, L., \ApJ  {\bf
521}, 179 - 189 (1999)
%  Title:  Aspherical Explosion Models for SN 1998BW/GRB-980425
%

\bibitem{Ho02}  H{\"o}flich, P., \etal, in {\it Proc. IAU Symposium 212 on
Massive Stars}, D. Reidel Conf. Series, ed. E. van den Hucht, (2002),
astro-ph/0207272
%  Title: Aspherical Supernovae Explosions
%

\bibitem{Hoe+01}  Hoerandel, J.R., \etal, {\it Proc. 27th ICRC, Hamburg},
{\bf 1}, 71 (2001)
%  KASCADE data
%

\bibitem{Hoe02} Hoerandel, J.R., \ApP  {\bf  } (in press)  (2002),
astro-ph/0210453
%  Title: On the knee in the energy spectrum of cosmic rays
%

\bibitem{Hunter97}  Hunter, S.~D., \etal, \ApJ  {481}, 205 (1997)
%  title = "{EGRET Observations of the Diffuse Gamma-Ray Emission from
%  the Galactic Plane}"
%

\bibitem{Ide+01}  Ide, Y., \etal, \PASJ  {\bf 53}, 1153 - 1162 (2001)
%  Title:  Propagation of Ultra-High Energy Cosmic Rays from Sources in
%  the Super-Galactic Plane.
%  from abstract:  "... this problem may be solved if we
%  assume that UHECRs come from some galaxies, such as AGNs and radio
%  galaxies."
%

\bibitem{Jokipii87} Jokipii, J.~R., \ApJ  {\bf 313}, 842--846 (1987)
% title = "{Rate of energy gain and maximum energy in diffusive shock
%  acceleration}"
%

\bibitem{QGSJET}  Kalmykov, N.N., \& Ostapchenko, S.S., {\it Yad.
Fiz.}  {\bf{56}}  105 (1993); {\it Phys.At.Nucl.} {\bf{56}}, 346 (1993)
%

\bibitem{Ka64}  Kardashev, N. S., {\it Astronomicheskii
  Zhurnal} {\bf 41}, 807 (1964)
%  Title :  Magnetic Collapse and the Nature of Intense Sources of Cosmic
%  Radio-Frequency Emission

\bibitem{Kronberg94}  Kronberg, P.P., {\it Rep.~Prog.~Phys.}, {\bf 57},
    325 - 382 (1994)
%  Title:  Extragalactic magnetic fields
%

\bibitem{Kr77}  Krymskii, G. F., {\it Akademiia Nauk SSSR, Doklady}  {\bf
234}, 1306 - 1308 (1977), in Russian
%  Title:  A regular mechanism for accelerating charged particles at the
%  shock front
%

\bibitem{Kulsrud99}  Kulsrud, R.M.,  \ARAA {\bf 37}, 37 (1999)
%  A Critical Review of Galactic Dynamos
%

\bibitem{Lagage+C83}  Lagage, P.~O., \& Cesarsky, C.~J., \AA {\bf 125},
249--257 (1983)
% title = "{The maximum energy of cosmic rays accelerated by supernova
%  shocks}",

\bibitem{LM2000} Learned, J.G. \& Mannheim, K., {\it Ann. Rev. Nucl.
 \& Part. Sci.} {\bf 50}, 679 - 749 (2000)
%  Title:  High energy neutrino astrophysics
%

\bibitem{MOD01}  Malkov, M. A., O'C Drury, L., {\it Rep. on Progr. in
Phys.} {\bf 64}, 429 - 481 (2001)
%  Title:  Nonlinear theory of diffusive acceleration of particles by
%  shock waves
%

\bibitem{Mezger96}   Mezger, P. G., Duschl, W. J., Zylka, R., \AAR  {\bf
7}, 289 - 388 (1996)
%  Title:  The Galactic Center: a laboratory for AGN?
%

\bibitem{MF2001}  Melia, F, Falcke, H., \ARAA {\bf 39}, p. 309-352 (2001)
% review on the GC
%

\bibitem%[Meyer, Drury, \& Ellison 1997]
{JPMeyer97} Meyer, J.-P., Drury,
L. O'C., Ellison, D. C.,   {\it ApJ}  {\bf 487}, 182  (1997)
%  Title:  Galactic Cosmic Rays from Supernova Remnants. I. A
%  Cosmic-Ray Composition Controlled by  Volatility and
%  Mass-to-Charge Ratio
%

\bibitem{MABK03}   Moiseenko, S.G., Ardeljan, N.V., Bisnovatyi-Kogan,
G.S.,   {\it Rev. Mex. Astron. \& Astroph., (Serie de Conferencias)},
{\bf 15}, 231 - 233 (2003)
%  Title:  Supernovae type II: Magnetorotational explosion
%

\bibitem{Mori97}  Mori, M., \ApJ  {\bf  478}, 225 (1997)
%  Title:  The Galactic Diffuse Gamma-Ray Spectrum from Cosmic-Ray Proton
%  Interactions
%

\bibitem{NW2000}  Nagano, M., Watson, A.A., {\it Rev. Mod.
Phys.},  {\bf 72}, 689 - 732 (2000).
%  Observations and implications of the ultra-high energy cosmic
%  rays
%

\bibitem{Na01}   Nakamura, T., Mazzali, P. A., Nomoto,
K., Iwamoto, K., \ApJ  {\bf  550}, 991 - 999 (2001)
%  Title:  Light Curve and Spectral Models for the Hypernova SN 1998BW
%  Associated with GRB 980425.
%  Explosion has more than $10^{52}$ erg, and is highly anisotropic
%

\bibitem{Pa98} Paczynski, B., in  {\it Proc. Gamma-Ray Burst}: 4th
Huntsville Symposium, Huntsville, AL Sept. 1997. Eds. by Ch. A.
Meegan \etal. Woodbury, New York :  AIP conference proceedings ;
{\bf 428}, p.783 (1998)
%  Title:  Gamma-Ray Bursts as Hypernovae
%

\bibitem%[Peters 1959]
{Peters59}  Peters, B., {\it Nuovo Cimento Suppl.}, {\bf XIV}, 436 - 456
(1959)
%  Title:  Origin of cosmic radiation
%

\bibitem%[Peters 1961]
{Peters61}  Peters, B., {\it Nuovo Cimento}, {\bf XXII}, 800 - 819
(1961)
%  Title:  Primary cosmic radiation and extensive airshowers
%

\bibitem{Piran99}  Piran, T., {\it Physics Reports} {\bf 314}, 575 (1999)
%  Title:  Gamma-ray bursts and the fireball model.
%

\bibitem{Pt99}  Ptuskin, V.S. \etal, ICRC 26, ms.  OG 3.2.32, vol. 4, p.
291 (1999)
%  Title:  The modified weighted slab technique:  Results
%  grammage as rigitidy $R$ given by $R^{-0.54}$ above 5 GV
%

\bibitem{GRB1}  Pugliese, G., Falcke, H., Biermann, P.L., \AAL { \bf 344},
    L37 - L40 (1999), astroph/9903036
%  Title:  A jet-disk symbiosis model for Gamma Ray Bursts:  SS 433
%  the next?
%

\bibitem{GRB2}   Pugliese, G., Falcke, H., Wang, Y., Biermann, P.L.,
\AA {\bf 358}, 409 - 416 (2000), astro-ph/0003025
%  Title:  The jet-disk model for GRBs:  cosmic ray and neutrino
%  backgrounds
%

\bibitem{UHECRI} Rachen, J.P., \& Biermann, P.L., \AA  {\bf 272},
161 (1993), astro-ph/9301010
%  Title:  Extragalactic ultra-high energy cosmic rays I.
%  Contribution from hot spots in FR-II radio galaxies
%

\bibitem{UHECRII}  Rachen, J.P., Stanev, T., \& Biermann, P.L., \AA
 {\bf 273}, 377 (1993), astro-ph/9302005
%  Title:  Extragalactic ultra high energy cosmic rays II.
%  Comparison with experimental data
%

\bibitem{RAMK94}  Ratkiewicz, R., Axford, W. I., \& McKenzie, J. F., \AA
{\bf  291}, 935 - 942 (1994)
%  Title:  Similarity solutions for synchrotron emission from a supernova
%  blast wave.
% discusses instability - but the argument follows Gary Zank's work
%

\bibitem{Sch02}  Schaefer, B.E., \etal, \ApJ (submitted) (2002),
astro-ph/0211189
% Title: GRB021004: a Massive Progenitor Star Surrounded by Shells
%

\bibitem{Schatz02}  Schatz, G., \ApP  {\bf  17},  13-22 (2002)
%  Title:  A search for fine structure of the knee in EAS size spectra
%

\bibitem%[Seemann \& Biermann 1997]
{SB97} Seemann, H. \& Biermann, P.L.,
\AA  {\bf 327}, 273 (1997), astro-ph/9706117.
%  Unstable waves in winds of magnetic massive stars
%

\bibitem{ST90}  Silberberg, R., \& Tsao, C. H., \ApJL  {\bf  352}, L49 -
L52 (1990)
%  Title:  An explanation for cosmic-ray source abundances including
%  nitrogen.
%  some influence from WR stars
%

\bibitem{ROSAT97} Snowden, S.L., \etal, {\em Astrophys. J.} {\bf
485}, 125 (1997)
%  ROSAT survey diffuse X-ray background, radial extent 5.6 kpc,
%  vertical scale height 1.9 kpc,
%  in plane electron density 0.0035 cm$^{-3}$, temperature $ 4 \, 106$ K.
%

\bibitem{CRIV} Stanev, T., Biermann, P.L., \& Gaisser, T.K.,
\AA {\bf 274}, 902 (1993), astro-ph/9303006
%  Title:  Cosmic rays IV. The spectrum and chemical composition above
%   $10^4$ GeV

\bibitem{Swordy}Swordy, S.P., Fortson, L.F., \etal, {\it Astropart.
Phys.} {\bf{18}}, 129 (2002), astro-ph/0202159
%

\bibitem{Ta+98}  Takeda, M., \etal, \PRL  {\bf 81}, 1163 - 1166 (1998)
%  Title:  Extension of the Cosmic-Ray Energy Spectrum beyond the
%  Predicted Greisen-Zatsepin-Kuz'min Cutoff.
%  AGASA data
%

\bibitem{Ta+99}  Takeda, M., \etal, \ApJ {\bf 522}, 225 - 237 (1999)
%  Title:  Small-Scale Anisotropy of Cosmic Rays above 10^19 eV Observed
%  with the Akeno Giant Air Shower Array.
%  Pairs and triplets
%

\bibitem{Samo}  Ter-Antonyan, S.V., \& Haroyan, L.S., {\it  Proc. of
the workshop  ANI98}, Forschungszentrum Karlsruhe, FZKA 6216, 115
(1998)
%

\bibitem{TH}  Ter-Antonyan, S.V.~, \& Haroyan, L.S., (2000) hep-ex/0003006
%

\bibitem{STPB} Ter-Antonyan, S.V., \& Biermann, P.L., astro-ph/0106076
(2001)
%

\bibitem{Ul+01}  Ulrich, H., \etal, {\it Proc. 27th ICRC, Hamburg}, {\bf
1}, 97 (2001)
%  KASCADE data
%  in STA-list as {KAS1}####################

\bibitem{HJV+PLB88}  V{\"o}lk, H.~J. \& Biermann, P.~L., \ApJL {\bf 333},
L65 - L68 (1988)
% title = "{Maximum energy of cosmic-ray particles accelerated by
%  supernova remnant shocks in stellar wind cavities}"

\bibitem{Wa02}  Wang, L., \etal, \ApJ  {\bf 579}, 671 - 677 (2002),
astro-ph/0205337
% Title: The Axially Symmetric Ejecta of Supernova 1987A
%

\bibitem{Wh02}  Wheeler, J.C., {\it Am.J.Phys.}  {\bf 71}, 11 - 22 (2003),
astro-ph/0209514
% Title: Observations and Theory of Supernovae
% Comments: AAPT/AJP Resource Letter
%

\bibitem{LB-CR}  Wiebel-Sooth, B., Biermann, P.L.,
   Landolt-B{\"o}rnstein, vol. VI/3c, Springer Verlag, p. 37 - 90 (1999)
%  general review on cosmic rays physics, including long lists of
%  experiments,
%  a section on UHECR experiments is by A.A. Watson
%

\bibitem{Yo+02a}  Yoshiguchi, H., Nagataki, S., Tsubaki, S., Sato, K.,
\ApJ  {\bf }, (in press) (2002),  astro-ph/0210132
%  Title:  Small Scale Clustering in Isotropic Arrival Distribution of
%  Ultra-High Energy Cosmic Rays and Implications for Their Source
%  Candidates
%

\bibitem{Yo+02b}  Yoshiguchi, H., \etal, \PASJ  {\bf }, (in press)
(2002),  astro-ph/0212061
%  Title:  Revision of the Selection Function of the Optical Redshift
%  Survey using the Sloan Digital Sky Survey Early Data Release: Toward an
%  Accurate Estimate of Source Number Density of Ultra-High Energy
%  Cosmic Rays
%

\end{thebibliography}
\end{document}